\documentclass[aps,pre,preprint,10pt,superscriptaddress,showpacs]{revtex4-1}
\usepackage{amsfonts} 
\usepackage{amsmath}
\usepackage{amssymb}
\usepackage{graphicx}
\usepackage{subfigure}
\usepackage{color}
\newcommand*\xbar[1]{%
  \hbox{%
    \vbox{%
      \hrule height 0.5pt 
      \kern0.5ex
      \hbox{%
        \kern-0.1em
        \ensuremath{#1}%
        \kern-0.1em
      }%
    }%
  }%
} 
\begin{document}
\title{Nonlinear  Landau damping and modulation  of electrostatic waves   in  a  nonextensive electron-positron  pair  plasma}
\author{Debjani Chatterjee}
\author{A. P. Misra}
\email{apmisra@visva-bharati.ac.in; apmisra@gmail.com}
\affiliation{Department of Mathematics, Siksha Bhavana, Visva-Bharati University, Santiniketan-731 235, West Bengal, India}

\begin{abstract}
The nonlinear theory of amplitude modulation of electrostatic wave envelopes in a collisionless electron-positron (EP) pair plasma is studied by using a set of  Vlasov-Poisson equations in the context of Tsallis' $q$-nonextensive statistics. In particular, the previous linear theory of Langmuir oscillations in EP plasmas [Phys. Rev. E {\bf87}, 053112 (2013)] is rectified and modified.  Applying the multiple scale technique (MST), it is shown that the evolution of electrostatic wave envelopes is governed by a nonlinear Schr{\"o}dinger (NLS) equation with a nonlocal nonlinear term  $\propto {\cal{P}}\int|\phi(\xi',\tau)|^2d\xi'\phi/(\xi-\xi') $  [where ${\cal P}$ denotes the Cauchy principal value, $\phi$ is the small-amplitude electrostatic (complex) potential, and $\xi$ and $\tau$ are the stretched coordinates in MST] which appears due to the wave-particle resonance.   It is found that a subregion $1/3<q\lesssim3/5$ of    superextensivity  $(q<1)$  exists  where   the carrier wave frequency can turn over with  the group velocity  going to zero and then to negative values.   The effects of the nonlocal nonlinear term and the nonextensive parameter $q$  are examined on the modulational instability (MI) of  wave envelopes as well as on the solitary wave solution of the NLS equation. It is found that the modulated wave packet is always unstable (nonlinear Landau damping) due to the nonlocal nonlinearity in the NLS equation.  Furthermore,  the effect of the nonlinear Landau damping  is to  slow  down the   amplitude of the wave envelope, and the corresponding decay rate can be faster the larger is the number of superthermal particles in pair plasmas. 

\end{abstract}
\maketitle
\section{Introduction} \label{sec-introduct}
Electron-positron (EP) pair plasmas are ubiquitous and play important roles in  many astrophysical situations such as   the early universe \cite{misner1973}, Van Allen radiation belts, and near the polar cap of fast rotating neutron stars \cite{lightman1982}, black holes \cite{blandford1977}, pulsars \cite{goldreich1969}, quasars \cite{wardle1998}, active galactic nuclei \cite{begelman1984}, accretion disks \cite{orsoz1997}  as well as in laboratories \cite{sarri2015}. In black holes, pulsars and quasars such EP plasmas are emitted in the form of ultra-relativistic winds or collimated jets by some of their most energetic objects. Because of the intrinsic and complete symmetry between the positively charged (e.g., positrons or positive ions) and negatively charged (e.g., electrons or negative ions) particles, the dynamics of pair plasmas become   significantly different from that of  electron-ion plasmas or from a purely electronic beam.  
Over the last few years, a number of experiments have been performed to create EP plasmas  (See, e.g., Refs. \onlinecite{surko1989,boehmer1995}). In such  experiments, it has been observed that because of the fast annihilation and the formation of positronium atoms, the identification of collective  modes in EP plasmas is practically impossible. To resolve this issue and to identify the collective modes properly,  a number of experiments have been proposed to create pair-ion plasmas (See, e.g., Ref. \onlinecite{oohara2003,oohara2005}).    However, most recently, ion free high-density ($\sim 10^{16}$ cm$^{-3}$) neutral EP plasmas with unique characteristics have been produced in the laboratory by using a compact laser driven setup \cite{sarri2015}. It has been reported that because of their unique characteristics together with the charge neutrality, small divergence as well as high average Lorentz factor, such EP plasmas can  exhibit collective behaviors and  thereby opening up  the possibility of studying the collective dynamics of EP plasmas in a controlled laboratory environment. 

 The collective oscillations of EP plasmas and associated wave dynamics together with the formation of solitary waves and shocks have been extensively studied over the past two decades (See, e.g., Refs. \onlinecite{jao2014,saberian2013,misra2004,liu2014}).  In other works, the modulational instability (MI) and the nonlinear evolution  of electrostatic and electromagnetic wave envelopes  have also been studied in EP plasmas (See, e.g., Refs. \onlinecite{mofiz1990,asenjo2012,javan2012}).  However, most of these works are based  on  hydrodynamic models.  Using the kinetic theory approach, i.e., using Vlasov-Poisson equations, the effects of Landau damping   on various kinds of wave modes (linear theory) as well as on  electrostatic and electromagnetic  solitary waves [governed by the Korteweg-de Vries (KdV) equation]  have also been studied in EP plasmas \cite{saberian2013} and in some other environments (See, e.g. Refs. \onlinecite{ott1969,misra2014,misra2015,brodin2015} and references therein).   However, to our knowledge, no theory for the explanation of electrostatic wave modulation and the effects of nonlinear Landau damping on MI and nonlinear evolution  have been reported in nonextensive EP plasmas. 

Many spacecraft observations (e.g., in the Earth's bow-shock,  upper ionosphere of Mars, the vicinity of the Moon etc.)  \cite{montgomery1968}  and laboratory experiments \cite{liu1994} confirm the presence of nonthermal and superthermal particles which do not follow the Maxwellian distribution, but show some deviation from the thermodynamic equilibrium. The presence of such nonthermal particles   has also been confirmed in astrophysical environments \cite{pierrard2010}. Several models for phase-space plasma distributions with nonthermal or superthermal wings or other deviations from Maxwellian distribution have been proposed in recent years. One of such distributions, which was first reported by Renyi \cite{renyi1955} and subsequently proposed by Tsallis \cite{tsallis1988}, is the Boltzmann-Gibbs-Shannon (BGS) entropy in which the degree of nonextensivity of the plasma particles is characterized by the entropic index $q$. The   distribution function with $q<1$ characterizes the system with more superthermal particles (superextensivity), while the distribution with $q>1$ indicates plasmas containing a large number low-speed particles (subextensivity)  compared to the Maxwellian one $(q\rightarrow1)$.  Such $q$-nonextensive distribution has been widely considered in a number of works to investigate various collective plasma modes and nonlinear coherent structures (See, e.g., Refs. \onlinecite{saberian2013,caruso2008,nobre2011,guo2013}). It is to be noted that the $\kappa$ distribution function and the $q$-nonextensive distribution function in Tsallis' statistics  are somewhat equivalent in the sense that in both these cases, the spectrum of the distribution functions show similar behaviors. In fact, there is a formal transformation $\kappa=1/(1-q)$ which can provide the missing links between these two velocity distribution functions \cite{leubner2002}
   
On the other hand, it is well known that waves in plasmas can undergo collisionless damping when they resonantly interact with   trapped and/or free particles, i.e., when the particle's velocity approaches  the wave phase velocity or group velocity.  Such collisionless damping was first theoretically   predicted by Landau \cite{landau1946}, and later confirmed experimentally by Malmberg {\it et al.} \cite{malmberg1964}.
Motivated by these inventions, Ott {\it et al.} \cite{ott1969} first theoretically investigated the effects of linear Landau damping on the nonlinear propagation of ion-acoustic solitons in electron-ion plasmas through the description of KdV equations and  on the assumption that particle's trapping time is much longer than that of Landau damping. Later,  Ikezi {\it et al.} \cite{ikezi1971}, based on  their experimental investigation,  have emphasized the importance of nonlinear Landau damping  in plasmas.  Accordingly,  Ichikawa \cite{ichikawa1974} investigated the  the effects of nonlinear Landau damping due to resonant particles having the group velocity of the wave, on the modulation of electrostatic wave envelopes in electron-ion plasmas. In this work, he assumed that the typical time scale is much longer than the bouncing period of particles trapped in the potential trough. It has been shown that the nonlinear wave-particle resonance leads to the modification of the nonlinear Schr{\"o}dinger (NLS) equation with a nonlocal nonlinear term proportional to a Cauchy principal value integral as well as the local (cubic) nonlinear term. Furthermore, in Ref. \onlinecite{ichikawa1974} it was reported that, in contrast to the ordinary NLS equation, the nonlinear resonance always leads to MI of wave envelopes against a plane wave perturbation regardless of the sign of $PQ$ positive or negative, where $P$ is the coefficient of the group velocity dispersion and $Q$ is the cubic nonlinear term of the NLS equation. 

In this work, our aim   is to revisit and extend the work of Ichikawa \cite{ichikawa1974} in an unmagnetized collisionless EP pair plasma in the context of  Tsallis' $q$-nonextensive  statistics. Starting from a set of Vlasov-Poisson equations  and using the reductive perturbation technique (RPT), we show that the NLS equation is  not only modified by the nonlinear resonant effects but also by the nonextensive parameter $q$ which contributes to both the dispersive and nonlinear (local and nonlocal) terms. It is found that the nonextensivity significantly modifies the wave frequency,   the group velocity,  the nonlinear frequency shift and the energy transfer rate for the modulated wave packets as well as the amplitude of solitary wave solutions of the NLS equation. 

\section{Basic Equations and derivation of NLS equation}
This section mainly focuses on the derivation of the NLS equation. Though, the relevant analysis is almost the same as in Ref. \onlinecite{ichikawa1974}, we, however, review the analysis and  present the subsequent derivations and explanations for the expressions of various plasma modes  in more details.    We consider the nonlinear  propagation of electrostatic wave packets along the $x$-axis in a collisionless electron-positron pair plasma.  The basic equations for the dynamics of EP plasmas  are given by the following Vlasov and Poisson equations 
\begin{equation}
\frac{\partial F_{\alpha}}{\partial t}+v\frac{\partial F_{\alpha}}{\partial x}-\frac{e_\alpha}{m_\alpha}\frac{\partial \phi}{\partial x} \frac{\partial F_{\alpha}}{\partial v}=0, \label{vlasov}
\end{equation}
\begin{equation}
 {\frac{\partial^2 \phi}{\partial x^2}}=-4\pi\sum e_\alpha \int F_\alpha dv, \label{poisson}
\end{equation}
where $F_\alpha$ is the distribution function with its unperturbed part $F^{(0)}_\alpha (v)$, $v$ is the particle velocity and $\phi$ is the electrostatic potential. Also,  the particle  charge and mass are given by  $ e_\alpha=-e$, $m_\alpha=m_e$ for electrons and $ e_\alpha=e$, $m_\alpha=m_p$ for positrons.
We assume that the equilibrium state (at $t=0$) is spatially uniform field-free EP plasma and  that the perturbation from the equilibrium state is purely electrostatic. Furthermore, we   consider the equilibrium distribution of electrons and positrons  $F^{(0)}_\alpha (v)$  to be the   $q$-distribution function as  in the Tsallis' nonextensive statistics \cite{tsallis1988}. It is to be mentioned that due to the resonance of plasma particles having the group velocity of the wave, the  distribution functions become singular, and so  the direct application of  RPT to the Vlasov-Poisson equations \eqref{vlasov} and \eqref{poisson} may not determine uniquely the contributions of the resonant particles. In order to treat these singularities properly   we introduce the multiple space-time scales as \cite{ichikawa1974}
\begin{equation}
   x\rightarrow x+\epsilon ^{-1} \eta +\epsilon^{-2}\zeta,~~  
  t\rightarrow t+\epsilon ^{-1}\sigma.    \label{stretch}
  \end{equation}
We assume that  the   amplitude of the carrier wave (with wave number $k$ and the wave frequency $\omega$) is infinitesimally small, and so  for $t>0$  a slight deviation $[\sim o(\epsilon)]$ from the uniform equilibrium  value will occur. Thus, we expand   
\begin{equation}
\begin{split}
&F_\alpha(v,x,t)= F^{(0)}_\alpha (v)+\sum_{n=1}^{\infty}\epsilon^{n} \sum_{l=-\infty}^{\infty}f^{(n)}_{\alpha,
l}(v,\eta,\sigma,\zeta)\exp{[il(kx-\omega t)]}  \\
&\phi(x,t)=\sum_{n=1}^{\infty}\epsilon^{n} \sum_{l=-\infty}^{\infty}\phi^{(n)}_l(\eta,\sigma,\zeta) \exp{[il(kx-\omega t)]},     \label{expansion}
\end{split}
\end{equation}
where $\epsilon$ is a small $(\lesssim1)$ positive parameter measuring the weakness of the wave amplitude and   $f^{(n)}_{\alpha,-l}=f^{(n)\ast}_{\alpha,l}$, $\phi^{(n)}_{-l}=\phi^{(n)\ast}_l$ for reality conditions to hold.  Here,   the asterisk denotes the complex conjugate quantity. We note that the expansion \eqref{expansion} for $F_{\alpha}$ is valid only in the non-resonance region where $|v-\omega/k|\gg o(\epsilon)$ is satisfied. In the resonance region where $v\approx \omega/k$, the above expansion may not be appropriate to apply to the Vlasov equations. So, the components $f^{(n)}_{\alpha,l}$ and $\phi^{(n)}_l$ are further expanded into multi-scale Fourier-Laplace integrals as
 \begin{equation}
 \begin{split}
 &f^{(n)}_{\alpha,l}(v,\eta,\sigma,\zeta)= \frac {1}{(2\pi)^2} \int_C d\Omega \int_ {-\infty}^{\infty}dK\tilde{f}^{(n)}_{\alpha,l}(v,K,\Omega,\zeta) \exp[i(K\eta-\Omega\sigma)] \\
& \phi^{(n)}_l(\eta,\sigma,\zeta)= \frac {1}{(2\pi)^2} \int_C d\Omega \int_{-\infty}^{\infty}dK\tilde{\phi}^{(n)}_l (K,\Omega,\zeta) \exp[i(K\eta-\Omega\sigma)], \label{Fourier-Lap-int}
\end{split}
 \end{equation}
where the contour $C $ is parallel to the real axis and lies above the coordinate of convergence.
 
We first derive the evolution equation of electrostatic wave envelopes for arbitrary species of particles with unperturbed arbitrary velocity distribution function, and then we extend our analysis in neutral EP plasmas in the context of Tsallis' $q$-nonextensive statistics \cite{tsallis1988}.  So, we substitute the stretched coordinates \eqref{stretch} and   the expansions \eqref{expansion} into Eqs. \eqref{vlasov} and \eqref{poisson} to obtain, respectively, as 
\begin{eqnarray}
&&il(\omega-kv)f^{(n)}_{\alpha,l}+ilkG_\alpha \phi^{(n)}_l\doteq\frac{\partial }{\partial \sigma}f^{(n-1)}_{\alpha,l}+v\frac{\partial }{\partial \eta}f^{(n-1)}_{\alpha,l}+v\frac{\partial }{\partial \zeta}f^{(n-2)}_{\alpha,l}-G_\alpha\frac{\partial }{\partial \eta}\phi^{(n-1)}_l-G_\alpha\frac{\partial }{\partial \zeta}\phi^{(n-2)}_l\notag\\
&&-ik\frac{e_\alpha}{m_\alpha}\sum^\infty_{s=1}\sum^\infty_{l'=-\infty}(l-l')\phi^{(n-s)}_{l-l'} \frac{\partial }{\partial v}f^{(s)}_{\alpha,l'}-\frac{e_\alpha}{m_\alpha}\sum^\infty_{s=1}\sum^\infty_{l'=-\infty}\left(\frac{\partial}{\partial \eta} \phi^{(n-s-1)}_{l-l'}+ \frac{\partial }{\partial \zeta}\phi^{n-s-2}_{l-l'}\right)  \frac{\partial }{\partial v}f^{(s)}_{\alpha,l'}, \label{vlasov1}
\end{eqnarray}
\begin{equation}
(lk)^2\phi^{(n)}_l-2ilk\frac{\partial}{\partial \eta}\phi^{(n-1)}_l -i2lk \frac{\partial }{\partial \zeta}\phi^{(n-2)}_l 
-\frac{\partial^2}{\partial \eta^2}\phi^{(n-2)}_l-4\pi\sum_{\alpha} e_\alpha\int f^{(n)}_{\alpha,l}dv=0.  \label{poisson1}
\end{equation}
where we have used the symbol $\doteq$   to denote the equality in the weak sense and   
disregarded the terms which contain $\phi^{(n-3)}_l$ and $\phi^{(n-4)}_l$ in Eq. \eqref{poisson1}.  

In the subsequent analysis, we  determine the contributions of the resonant particles having the wave group velocity   by solving the $\sigma$-evolution of the components   $f^{(n)}_{\alpha,l}$ and $\phi^{(n)}_l$ as an initial value problem  with the  initial condition
\begin{equation}
f^{(n)}_{\alpha,0} (v, \eta, \sigma=0, \zeta)\doteq0,~~~ n\geq 1, \label{int-cond1}
\end{equation}
in the multiple space-time scheme corresponding to that on the distribution function
\begin{equation}
f^{(n)}_{\alpha,0} (v, t=0)=0. \label{int-cond2}
\end{equation}

 \subsection{Harmonic modes for $n=l=1$: Linear dispersion law}
 Equating the coefficients of $\epsilon$ from Eqs. \eqref{vlasov1} and \eqref{poisson1} for $n=1,~l=1$,  we  obtain (For details see Appendix \ref{appendix1})  the following linear dispersion law:
 \begin{equation}
 k+4\pi\sum e_\alpha\int_C\frac{G_\alpha(v)}{\omega-kv}dv=0, \label{dispersion}
\end{equation}
together with the linear Landau damping rate given by 
\begin{equation}
 \gamma_L=\frac{\pi}{k}\sum_{\alpha}e_{\alpha}G_{\alpha}\left(\frac{\omega}{k}\right)/\sum_{\alpha}e_{\alpha}\int_C\frac{G_{\alpha}}{\left(\omega-kv\right)^2}dv, \label{dispersion-img}
 \end{equation}
 where   $\int_C$ denotes the Cauchy Principal value, and  we have made the analytical continuation of the integral over $v$  along the real axis passing infinitesimally above and under the pole at $v=\omega/k$ with the constraint of weakly damped waves and 
\begin{equation}
G_\alpha (v)=\frac{e_\alpha}{m_\alpha}\frac {\partial}{\partial v}F^{(0)}_\alpha (v).\label{G-alpha}
\end{equation}
\subsection{Harmonic modes with $l\neq 0,~n=1$: Some conditions}
From Eqs. \eqref{vlasov1} and \eqref{poisson1}, we equate the components for $l\neq 0$ and $n=1$, and then  use the dispersion relation \eqref{dispersion} to obtain the following conditions (See Appendix \ref{appendix2})
 \begin{equation}
 f^{(1)}_{\alpha,l}\doteq 0~\text{and}~\phi^{(1)}_l=0~\text{for}~|l|\geq 2. \label{cond-f-phi-l-1}
 \end{equation}
\subsection{Zeroth harmonic modes for $n=1,~2;~l=0$}
Here, we examine the second order terms with    $n=2$ and $l=0$. Thus, we have from Eq. \eqref{vlasov1}
 \begin{equation}
 \frac{\partial}{\partial \sigma}f^{(1)}_{\alpha,0}+v\frac{\partial}{\partial \eta}f^{(1)}_{\alpha,0} -G_\alpha\frac{\partial}{\partial \eta} \phi^{(1)}_0\doteq0, \label{vlasov-n2-l0}
 \end{equation}
 while the first order terms  with  $n=1$ and $l=0$ of Eq. \eqref{poisson1} yields
 \begin{equation}
 \sum e_\alpha \int f^{(1)}_{\alpha,0}dv=0. \label{poisson-n1-l0}
 \end{equation}
 Substituting the Fourier-Laplace integrals given by Eq. \eqref{Fourier-Lap-int} into Eq. \eqref{vlasov-n2-l0} and solving it as an initial value problem we obtain
  \begin{equation}
  \tilde{f}^{(1)}_{\alpha,0}(v,K,\Omega,\zeta)\doteq-\frac{i}{\Omega -Kv}f^{(1)}_{\alpha,0}(v,K,\sigma=0,\zeta)-\frac{K}{\Omega-Kv}G_\alpha  \tilde{\phi}^{(1)}_0 (K,\Omega,\zeta), \label{26}
   \end{equation}
where we have used an arbitrary constant as $-f^{(1)}_{\alpha,0}\left(v,K,\sigma=0,\zeta\right)$. 

Next,  a substitution of Eq. \eqref{26} into Eq. \eqref{poisson-n1-l0} gives
 \begin{equation}
  \tilde{\phi}^{(1)}_0 (K,\Omega,\zeta)=-{i\sum e_\alpha \int \frac{f^{(1)}_{\alpha ,0} (v,K, \sigma=0, \zeta)}{\Omega -Kv}dv}/{\sum e_\alpha \int \frac{G_\alpha}{\Omega -Kv}dv}. \label{27}
 \end{equation}
Using the initial condition \eqref{int-cond1} and taking Fourier inversion of Eqs. \eqref{26} and \eqref{27} we obtain for $n=1,~l=0$ the following zeroth order components:
 \begin{equation}
f^{(1)}_{\alpha,0}\doteq 0,~~~\phi^{(1)}_0=0.  \label{f-phi-n1-l0}
\end{equation}
The vanishing of $f^{(1)}_{\alpha,0}$ and hence of $\phi^{(1)}_0$ in fact eliminates the contributions of the modes associated with the singularities of the integrals in Eq. \eqref{27}. Thus, in this way the initial condition \eqref{int-cond1} uniquely defines the present problem  to investigate nonlinear auto modulation of $(\omega,k)$ modes.
 \subsection{Modes with $n=2,~l=1$: Compatibility condition}
We proceed to examine the  the second order, first harmonic modes with $n=2$ and $l=1$. Thus,   from Eqs. \eqref{vlasov1} and \eqref{poisson1} we obtain   the following compatibility condition  (See for details Appendix \ref{appendix3}) 
 \begin{equation}
\left\{ \frac{\partial}{\partial \sigma} + \lambda \frac{\partial}{\partial \eta}\right\} \phi^{(1)}_1(\eta, \sigma;\zeta) =0, \label{vg-eq}
\end{equation}
where $\lambda$ is  given by
\begin{equation}
\lambda=\left[4\pi\sum e_\alpha \int_C\frac{G_\alpha}{(\omega-kv)^2} dv\right]^{-1} \left[1+4\pi\sum e_\alpha \int_C\frac{v}{(\omega-kv)^2}G_\alpha dv\right], \label{lambda}
\end{equation}
We find that this expression of $\lambda$ is exactly the same as the group velocity  $v_g\equiv{\partial \omega}/{\partial k}$ obtained by differentiating the dispersion relation \eqref{dispersion} with respect to $k$.
Equation \eqref{vg-eq} shows that the $\sigma-\eta$ variation of $\phi^{(1)}_1$ can be related to a new coordinate  defined by
\begin{equation}
\xi=\eta-\lambda \sigma= \epsilon(x-\lambda t), \label{xi-transf}
\end{equation}
such that
\begin{equation}
\phi^{(1)}_1 (\eta, \sigma;\zeta)=\phi^{(1)}_1 (\xi;\zeta). \label{phi-xi-zeta}
\end{equation}
This indicates that the  coordinate $\xi$ in Eq. \eqref{xi-transf} establishes a clear relationship between the reductive perturbation theory and the multiple space-time expansion method.
\subsection{Second  harmonic modes with $n=l=2$}
For the second order quantities with $n=l=2$, we obtain  from  Eqs. \eqref{vlasov1} and \eqref{poisson1} the following expressions (See Appendix \ref{appendix4})
\begin{equation}
 f^{(2)}_{\alpha,2}\doteq - \frac{k}{\omega-kv+i\nu}\left[G_\alpha \phi^{(2)}_2 -\frac{e_\alpha}{2 m_\alpha}k \frac{\partial}{\partial v}\left(\frac{G_\alpha}{\omega-kv+i\nu}\right) \left(\phi^{(1)}_1\right)^2\right], \label{f-alpha2-n2-l2}
\end{equation}
\begin{equation}
\phi^{(2)}_2=\frac{1}{6} A(k,\omega)\left(\phi^{(1)}_1\right)^2, \label{phi2-n2-l2}
\end{equation}
where
\begin{equation}
A(k,\omega)=4\pi\sum_\alpha \frac{e_\alpha ^2}{m_\alpha}\int_C \frac{1}{\omega-kv}\frac{\partial}{\partial v}\left(\frac{G_\alpha}{\omega-kv}\right)dv. \label{A-k-omega}
\end{equation}
Note that in Eq. \eqref{A-k-omega} the contributions of the resonance terms at the phase velocity $\omega/k$ are neglected because our basic assumption [Eq. \eqref{dispersion-img}] for EP plasma oscillations is that the linear Landau damping is a higher-order effect than second order. 
\subsection{Modes with $n=3,~l=0$}
From Eqs. \eqref{vlasov1} and \eqref{poisson1} we consider   the terms corresponding to $n=3$, $l=0$, and use  the relations \eqref{f-phi-n1-l0}  and \eqref{vg-eq} to obtain a set of reduced equations which, after use of the Fourier-Laplace transforms   with respect to $\eta$ and $\sigma$ and   the  initial condition \eqref{int-cond1}, yield 
\begin{equation}
\tilde{f}^{(2)}_{\alpha,0}\doteq k^2\left[-\frac{{\cal{W}}(K,\Omega)}{\Delta^{(c)}(K,\Omega)} \frac{K}{\Omega-Kv}G_\alpha(v)-\frac{e_\alpha}{m_\alpha}\frac{K}{\Omega-Kv}I_\alpha(v)\right] H(K,\Omega), \label{fF-alpha-n2-l0-final}  
\end{equation}
\begin{equation}
\tilde{\phi}^{(2)}_0=k^2\frac{H(K,\Omega)}{\Delta^{(c)}(K,\Omega)}{\cal{W}} (K,\Omega), \label{phiF-n2-l0}
\end{equation}
where the relevant details and the expressions for $H(K,\Omega),~{\cal{W}}(K,\Omega),~\Delta^{(c)}(K,\Omega)$ and $I_\alpha(v)$ are given in Appendix \ref{appendix5}.
\subsection{Third order harmonic modes with $n=3,~l=1$: The NLS equation} \label{section-nls-equation}
Considering the terms for $n=3$ and $l=1$, we obtain from Eqs. \eqref{vlasov1} and \eqref{poisson1}   a set of reduced equations, which after few steps (See Appendix \ref{appendix6}), result  
  into   the following modified nonlinear Schr{\"o}dinger equation for the small but finite  amplitude perturbation $\phi(\xi, \tau)\equiv\phi^{(1)}_1(\xi, \tau)$ as 
\begin{equation}
i\frac{\partial\phi}{\partial \tau}+P \frac{\partial^2\phi}{\partial \xi^2}+Q |\phi|^2 \phi +\frac{R }{\pi}{\cal P}\int\frac{ |\phi(\xi',\tau)|^2}{\xi-\xi'} d\xi' \phi+iS\phi=0. \label{nls}
\end{equation}
 The coefficients of the dispersion (group velocity), cubic nonlinear (local, nonlocal nonlinear terms, respectively, are  $P,~Q$ and $R$, given by $P\equiv(1/2)\partial^2\omega/\partial k^2=\beta/ \alpha,~Q= \gamma/ \alpha$ and $R=\delta/ \alpha$, where  
\begin{equation}
\alpha=4\pi \sum_\alpha e_\alpha \int \frac{G_\alpha}{(\omega-kv)^2}dv, \label{alpha}
\end{equation}
\begin{equation}
\beta=4\pi \sum_\alpha e_\alpha \int \frac{(v-\lambda)^2}{(\omega-kv)^3}G_\alpha dv, \label{beta}
\end{equation}
\begin{equation}
\gamma=\left(\frac{1}{6}\frac{A^2}{k}+\frac{1}{2}B\right)k^2-\Theta(k,~ \omega), \label{gamma}
\end{equation}
\begin{equation}
\delta=-\Phi(k, \omega), \label{delta}
\end{equation}
with
\begin{equation}
B(k, \omega)=4\pi \sum_\alpha \frac{e^3_\alpha}{m^2_\alpha} \int \frac{1}{\omega-kv} \frac{\partial}{\partial v} \left[ \frac{1}{\omega-kv} \frac{\partial}{\partial v}\left( \frac{G_\alpha}{\omega-kv}\right) \right] dv. \label{B}
\end{equation}
 Furthermore, the coefficient $S$ of Eq. \eqref{nls}, representing the linear Landau damping rate associated with the resonant particles having the phase velocity of the carrier wave, is given by
\begin{equation}
S=\frac{\theta (p)\gamma_L}{\epsilon^2}. \label{S}
\end{equation}
In Eq. \eqref{nls}, the coefficient $P$ appears due to the group velocity dispersion of the wave envelope. The most significant contribution of the resonant particles having the wave group velocity is the appearance of the nonlocal nonlinear term $\propto R$. This resonance contribution also modifies  the local nonlinear coefficient $Q$, which  appears due to carrier wave self-interactions originating from the zeroth harmonic modes (or slow modes).
\section{Conservation laws} \label{sec-conservation-laws}
Before we proceed to the modulation of electrostatic wave envelopes in $q$-nonextensive EP plasmas, we  verify some important conservation laws that are associated with the ordinary NLS equation (Here, ordinary means in absence of any nonlocal nonlinearity in the NLS equation). We will show that the nonlocal nonlinear term in Eq. \eqref{nls} violates these conservation laws. We note that in absence of the Landau damping effects, i.e., for $R=S=0$, the NLS equation \eqref{nls} possesses an infinite number of conservation laws. The first three conserving quantities are namely, the mass $I_1=\int|\phi|^2d\xi$, the momentum $I_2=(2i)^{-1}\int\left(\phi^{\ast}\partial_{\xi}\phi-\phi\partial_{\xi}\phi^{\ast}\right)d\xi$ and the energy $I_3=\int\left(|\partial_{\xi}\phi|^2-(Q/2P)|\phi|^4\right)d\xi$. The similar quantities, however, for the NLS equation \eqref{nls} with Landau damping satisfy the following equations:
\begin{equation}
\left(\frac{\partial}{\partial\tau}+2S\right)I_1=0, \label{mass-conserv}
\end{equation}
\begin{equation}
\left(\frac{\partial}{\partial\tau}+2S\right)I_2+\frac{R}{\pi}{\cal P}\int\int \frac{1}{\xi-\xi'}|\phi(\xi',\tau)|^2\frac{\partial}{\partial\xi}|\phi(\xi,\tau)|^2 d\xi d\xi'=0, \label{momentum-conserv}
\end{equation}
\begin{equation}
\left(\frac{\partial}{\partial\tau}+2S\right)I_3+i\frac{R}{\pi}{\cal P}\int\int \frac{1}{\xi-\xi'}|\phi(\xi',\tau)|^2\frac{\partial}{\partial\xi}\left(\phi\frac{\partial^2}{\partial\xi^2}\phi^{\ast}-\phi^{\ast}\frac{\partial^2}{\partial\xi^2}\phi\right) d\xi d\xi'=0, \label{energy-conserv}
\end{equation}
where the terms $\propto S$ and $R$ (the Cauchy principal value integrals) are due to the linear and nonlinear Landau damping effects. Next,  in Eq. \eqref{energy-conserv}   using the fact that the integral over $\xi$ is a convolution  of the functions ${\cal P}\left[1/(\xi'-\xi)\right]$ and $\partial_{\xi}\varphi(\xi,\tau)$, where $\phi\partial_{\xi}^2\phi^{\ast}-\phi^{\ast}\partial_{\xi}^2\phi= \partial_{\xi}\left(\phi\partial_{\xi}\phi^{\ast}-\phi^{\ast}\partial_{\xi}\phi\right)\equiv \partial_{\xi}\varphi(\xi,\tau)$, and noting that the Fourier inverse transform of $i~\text{sgn}{(s)}=-(1/\pi){\cal P}(1/\xi)$, we have 
\begin{equation}
\int \frac{\partial \varphi(\xi,\tau)}{\partial\xi}{\cal P}\frac{1}{\xi'-\xi}d\xi=\frac{1}{2}\int\exp(is\xi')|s|\hat{\varphi}(s,\tau)ds.
\end{equation}
So, performing the integral over $\xi'$ as a Fourier transform of $|\phi(\xi',\tau)|^2$  we obtain
\begin{equation}
{\cal P}\int\int \frac{1}{\xi-\xi'}|\phi(\xi',\tau)|^2\frac{\partial \varphi(\xi,\tau)}{\partial\xi}d\xi d\xi'=\frac{1}{2}\int|s|\hat{\varphi}(s,\tau)|\hat{\phi}(-s,\tau)|^2ds,
\end{equation}
where `hat' denotes the Fourier transform with respect to $\xi$ or $\xi'$.
Furthermore, using  $\hat{\varphi}(s,\tau)\equiv-2is|\hat{\phi}(s,\tau)|^2$ we obtain from Eq. \eqref{energy-conserv} 
\begin{equation}
\left(\frac{\partial}{\partial\tau}+2S\right)I_3=-\frac{R}{\pi}\int s^2|\hat{\phi}(s,\tau)|^2|\hat{\phi}(-s,\tau)|^2ds. \label{energy-conserv-reduced}
\end{equation}
From this equation we observe that   if the linear  Landau damping  is   a higher-order effect than $\epsilon^2$, the term $\propto S$ can be neglected, and so the the left-hand side of Eq. \eqref{energy-conserv-reduced}  represents the rate of change   of the wave energy. Also, the integral on the right-hand side is positive definite, and to be shown later that  for long wavelength EP plasma oscillations,  $R$ is always positive for $1/3<q<1$ and $q>1$, and  negative for $q<1/3$. Furthermore, it has been shown in Ref. \onlinecite{saberian2013} and will be shown in this work also that  the relevant results may not be valid for $q<1/3$.   Thus,   in both the superextensive and subextensive EP plasmas we have the inequality (the equality holds for $\phi=0~\forall~\xi$) 
\begin{equation}
\frac{\partial I_3}{\partial\tau}\leq0, \label{dI2-dtau}
\end{equation} 
implying that an initial perturbation (e.g., in the form a soliton) will decay to zero with time $\tau$, and hence a steady state solution with $|I_3|<\infty$ of the NLS equation \eqref{nls} may not exist in presence of the nonlinear Landau damping term in nonextensive EP plasmas. 
\section{Electrostatic envelopes with nonextensive stationary state} \label{sec-langmuir-envelopes}
The NLS equation \eqref{nls}  in Sec. \ref{section-nls-equation} has been derived in a general way to describe the evolution of electrostatic waves in plasmas with  arbitrary  species $\alpha$ and with arbitrary velocity distribution function for equilibrium plasma state. However, the main purpose of the present work is to investigate the dispersion properties of carrier wave modes, the linear Landau damping rate associated with the resonant particles with the phase velocity,    the modulational instability and nonlinear Landau damping due to resonant particles having the group velocity as well as  nonlinear evolution of wave envelopes in  EP plasmas with $q$-nonextensive stationary states. In the latter,    the $q$-distribution function in one space dimension is given by \cite{ silva1998,curado1999,lima2001}
\begin{equation}
 F^{(0)}_\alpha (v)=A_{\alpha, q}\left[ 1-(q_\alpha-1)\frac{m_\alpha v^2}{2k_BT_\alpha}\right] ^{1/(q_\alpha-1)} , \label{q-distribution}
 \end{equation}
 where $m_\alpha$ and $T_\alpha$ are respectively the mass and temperature of $\alpha$-species particles  ($\alpha=e,~p$, respectively, stand for electrons and positrons) and $k_B$ is the   Boltzmann constant. The spectral index $q_\alpha$ gives a measure that determines the slope of the energy spectrum of nonthermal particles. It also measures the deviation from the thermal equilibrium state $(q_\alpha\rightarrow1)$. We consider a fully symmetric and charge-neutral  EP plasma in which $T_e=T_p=T$ and $m_e=m_p=m$. Such assumptions are justified with the experimental works in pure pair plasmas with particles having the same dynamics \cite{oohara2003,oohara2005}. Also, in the creation of a pure EP plasma, the whole system reaches  a common thermal state with equal particle temperature. So,   the spectral index for electrons and positrons are taken as the same, i.e., $q_e=q_p=q$ \cite{saberian2013}.   The normalization constant $A_{\alpha, q}$ is given by $A_{\alpha, q}=L_q\sqrt{{m_\alpha}/{2\pi k_B T_\alpha}}=L_q/\sqrt{2\pi} v_t$, where $v_t=\sqrt{k_B T/m}$ is the particle's thermal velocity and   $L_q$ is defined by
\begin{equation}
L_q=\left\lbrace\begin{array}{cc}1, &  q\rightarrow1,\\ \frac{\Gamma\left(\frac{1}{1-q}\right) } {\Gamma\left( \frac{1}{1-q}-\frac{1}{2}\right) }\sqrt{1-q},&   -1<q<1,\\\\
\left( \frac{1+q}{2}\right)  \frac{\Gamma \left( \frac{1}{2}+ \frac{1}{q-1}\right) }{\Gamma\left( \frac{1}{q-1}\right) }\sqrt{q-1},&   q>1. \end{array}\right.  \label{qltgt1}
\end{equation}
Note that in the extensive limit   $q\rightarrow1$, the distribution function \eqref{q-distribution} reduces to that of the standard Maxwell-Boltzmann distribution $F^{(0)}_{\alpha}(v)=\sqrt{{m_\alpha}/{2\pi k_B T_\alpha}}\exp\left(-m_\alpha v^2/2k_B T_\alpha\right)$.  However, in the present work the case of $q\rightarrow1$ may not be recovered directly because of    simplifications  of  some expressions in the subsequent analysis, e.g., for long-wavelength perturbations.    The distribution function with $q<1$ represents the superextensivity, i.e., presence of more (compared to the Maxwellian) particles with  velocities larger than their thermal velocities (superthermal particles), whereas the case of subextensive distribution is represented by $q>1$, meaning that there is a large number of particles with velocities lower than their thermal velocities. Furthermore, because of long-range nature of Coulomb forces between the plasma particles and the presence of many superthermal particles   in astrophysical environments   \cite{pierrard2010}, a $q$-distribution with $q<1$ is strongly suggested in real plasma systems or superthermal plasmas. For more about the behaviors of the distribution function $F^{(0)}_\alpha (v)$ for different values of  $q$, the readers are  referred to, e.g.,  Ref. \onlinecite{saberian2013}.  

In what follows, we consider the nonlinear propagation of high-frequency $(\omega>\omega_p)$ and long wavelength [$k\ll k_d$, where $k_d=\left(8\pi n_0 e^2/k_B T\right)^{1/2}$ is the Debye wave number and $n_{e0}=n_{p0}=n_0$ is the equilibrium plasma number density] oscillations in $q$-nonextensive plasmas  whose phase velocity greatly exceeds  the  thermal velocities of electrons and positrons, i.e.,  ${\omega_r}/{k}\gg v_{t}>v$. In this case, the linear dispersion laws and the coefficients of the modified NLS equation \eqref{nls} can be simplified. To this end, we evaluate the Cauchy Principal value integral in Eq. \eqref{dielectric real fnc} with the following expansion for the integrand \cite{saberian2013} 
\begin{eqnarray}
&&\int_{-v_\text{max}}^{v_\text{max}} \frac{\frac{\partial}{\partial v} \left(F^{(0)}_e (v)+F^{(0)}_p (v)\right)}{(\omega_r/k)-v}dv\notag\\
&&= \frac{k}{\omega_r} \int_{-v_\text{max}}^{v_\text{max}} \left( \frac{\partial F^{(0)}_e (v)}{\partial v} +\frac{\partial F^{(0)}_p (v)}{\partial v}\right)  \left( 1+ \frac{k}{\omega_r} v + \frac{k^2}{{\omega_r}^2} v^2 + \frac{k^3}{{\omega_r}^3} v^3 +\cdots\right) dv, \label{integration-expansion} 
\end{eqnarray}
where the integration limits are taken as $\pm v_\text{max}=\pm\infty$   for $q<1$ and $\pm\sqrt{{2k_BT}/{m(q-1)}}$  for $q>1$. These limits are considered due to the fact that for $q>1$, the distribution function \eqref{q-distribution} has a thermal cutoff, which limits the velocity of particles to a maximum value, i.e., $v<v_\text{max}$. However, such cutoff is absent when $q<1$. In this case, the velocity of particles remains unbounded (For details, see, e.g.,  Ref. \onlinecite{saberian2013}).  

Next,  to evaluate the integrals in Eq. \eqref{integration-expansion} we note  that the $q$-distribution function  $F^{(0)}_\alpha(v)$ is an even function of  $v$, while   ${\partial F^{(0)}_\alpha (v)}/{\partial v}$ is an odd function of  $v$. Thus, one can evaluate the integrals as follows:
\begin{equation}
\begin{split}
&\int_{-v_\text{max}}^{v_\text{max}} v^m\frac{\partial F^{(0)}_\alpha (v)}{\partial v} dv =0,~~\text{for}~m=0,2,4, \\
&\int_{-v_\text{max}}^{v_\text{max}} v \frac{\partial F^{(0)}_\alpha (v)}{\partial v} dv = -1,~~  
\int_{-v_\text{max}}^{v_\text{max}} v^3\frac{\partial F^{(0)}_\alpha (v)}{\partial v} dv = - \frac{6 v_{t}^2}{3q-1},\\
& \int_{-v_\text{max}}^{v_\text{max}} v^5\frac{\partial F^{(0)}_\alpha (v)}{\partial v} dv = - \frac{60 v_{t}^4}{(3q-1)(5q-3)}.   \label{integrals-v}
\end{split}
\end{equation}
We have evaluated the above integrals by parts and the average value of $v^2$ as
\begin{equation}
\langle v^2\rangle\equiv\int_{-v_\text{max}}^{v_\text{max}} v^2  F^{(0)}_\alpha (v)dv=\frac{2 v_{t}^2}{3q-1}. \label{average-v}
\end{equation}
We mention that for $q>1$, the above integrals, in which the limits are $\pm v_\text{max}$, are obtained by reducing the integrals to beta functions of the form $B(m,n)$ with $m,~n>0$ and finally to gamma functions using the relation $B(m,n)=\Gamma(m)\Gamma(n)/\Gamma(m+n)$. Similar integrals can also be evaluated for $-1<q<1$ in which the limits are $\pm\infty$ by using the relation  $B(m,n)=\int_0^{\infty}\frac{x^{n-1}}{(1+x)^{m+n}}dx$ and the above relation between beta and gamma functions. However, in each of these cases of superextensive and subextensive plasmas, we will obtain the same results by means of Eq. \eqref{qltgt1} except the factors $\sqrt{1-q}$  for $-1<q<1$ and  $\sqrt{q-1}$ for $q>1$. 
From Eq. \eqref{average-v} we also note that the values of $q~(<1)$ are further restricted to the region $1/3<q<1$, because otherwise, the average value of $v^2$ may diverge. In particular, in the limit $q\rightarrow1$, Eq. \eqref{average-v} reduces to the well known energy equipartition relation $\langle\frac{1}{2}m v^2\rangle=\frac{1}{2}k_BT$. Thus, our results may be  valid for both the superextensive $(1/3<q<1)$ and subextensive $(q>1)$ distributions of electrons and positrons in plasmas.

\begin{figure}[ht]
\centering
\includegraphics[height=2.5in,width=7in]{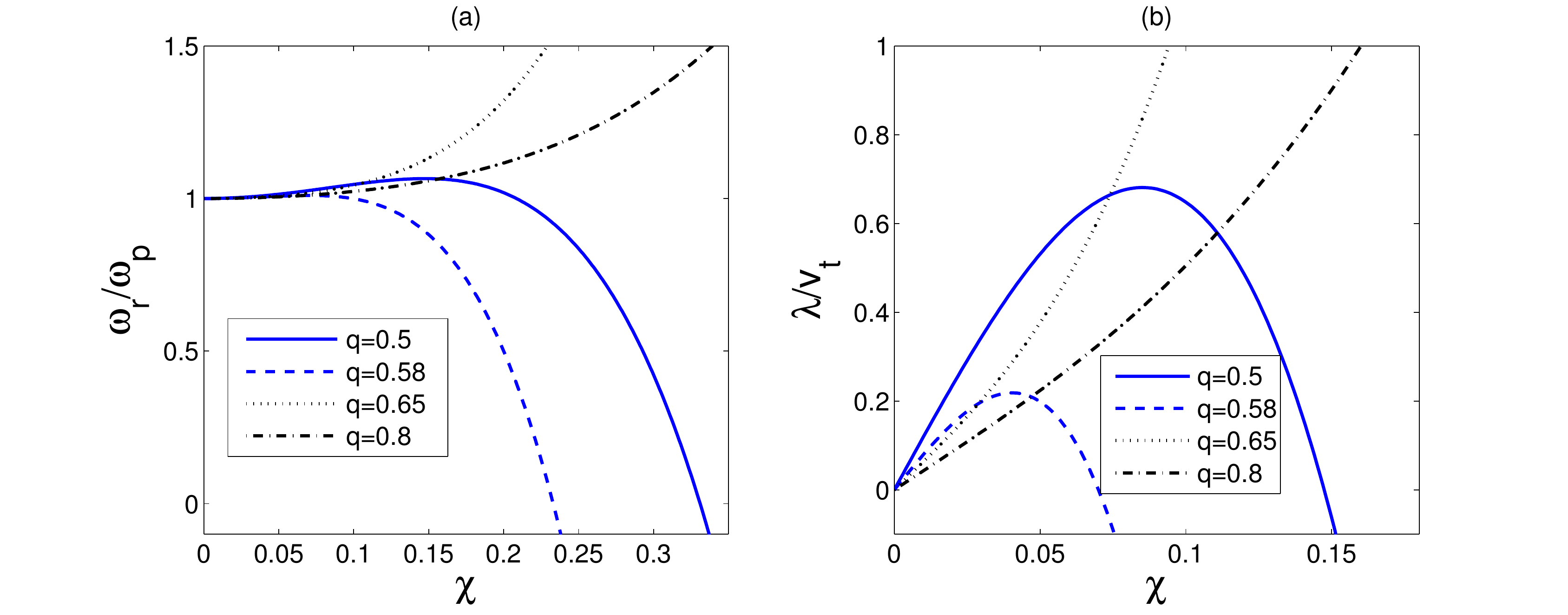}
\caption{ (Color online) The real part of the carrier wave frequency $\omega_r/\omega_p$ [Subplot (a); Eq. \eqref{dispersion-reduced}] and the group velocity $\lambda/v_t$ [Subplot (b); Eq. \eqref{lambda-grop-velocity}] of the wave envelope are plotted against the carrier wave number $\chi\equiv k/k_d$ for different values of the nonextensive parameters as in the figure. It is seen that the carrier  wave dispersion curve can turn over with the group velocity of the wave envelope going to zero and then to negative values (See the solid and dashed lines) in the subregion $0.47\lesssim q<3/5$ of the superextensive region $1/3<q<1$. In the region $1/3<1\lesssim0.47$, the values of $\lambda/v_t$ greater than unity are inadmissible, while in the other regime $q>3/5$, both the frequency and the group velocity assume only the positive values. }
\label{fig1}
\end{figure}

\begin{figure}[ht]
\centering
\includegraphics[height=2.5in,width=7in]{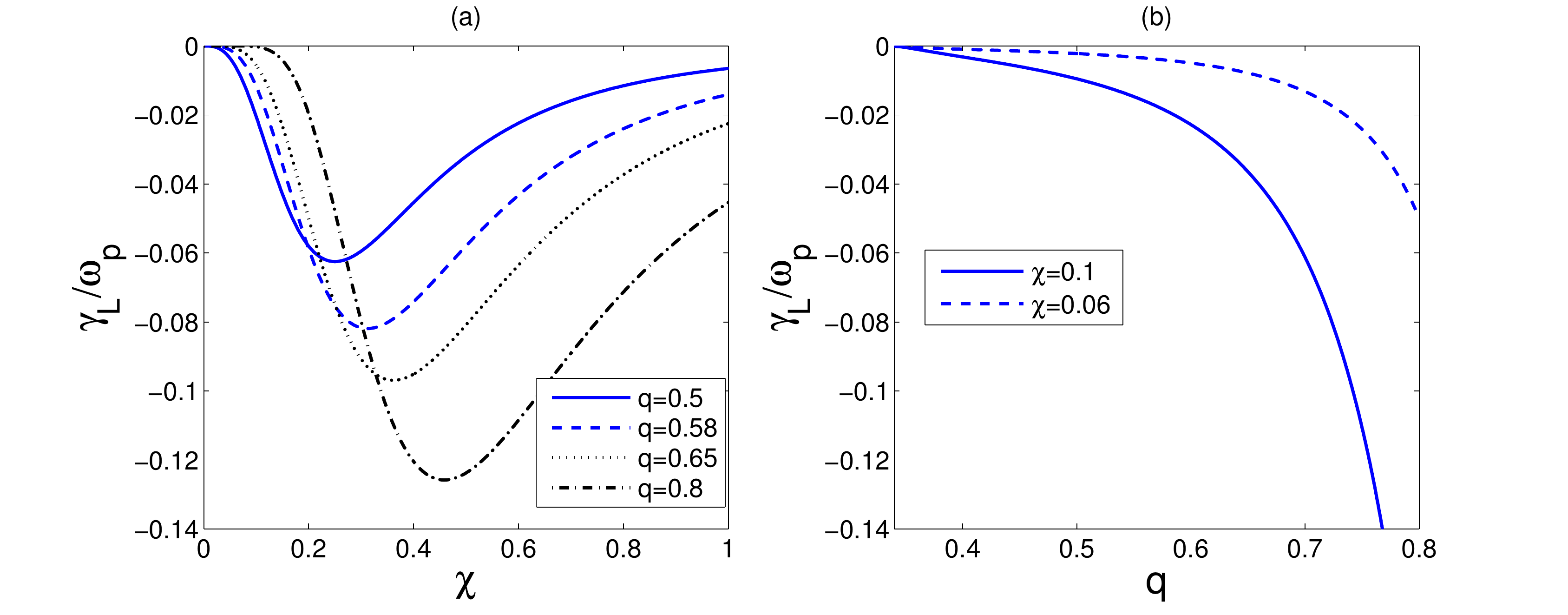}
\caption{(Color online) The imaginary part of the carrier wave frequency $\gamma_L/\omega_p$ [The linear Landau damping rate given by Eq. \eqref{landau-damping-linear-nonextensive}] is plotted against $\chi\equiv k/k_d$ [left panel (a)] and the nonextensive parameter $q$ [right panel (b)] for different values of $q$ and $\chi$ as in the figure. From the left panel it is seen that in one subregion $0\lesssim\chi\lesssim\chi_0$, the damping rate increases with $\chi$, while in the other $\chi_0\lesssim\chi\lesssim1$ it reduces with $\chi$. Furthermore, lower the percentage of superthermal particles, the higher is the Landau damping rate. The right panel shows that the damping rate becomes weaker in the limit of long-wavelength $(\chi\ll1)$ oscillations.  }
\label{fig2}
\end{figure}

 Now, we use the results as in Eqs. \eqref{integrals-v} and \eqref{average-v}, and in the region of small wave numbers $(1\gg\chi^2\equiv k^2/k_d^2)$ to  obtain from Eq. \eqref{dielectric real fnc} the following dispersion relation for electrostatic carrier waves in a nonextensive EP plasma: 
\begin{equation}
{\omega_r}^2 = {\omega_p}^2\left[ 1+ 3\chi^2 \frac{2}{3q-1} +60 \chi^4 \frac{1}{(3q-1)(5q-3)}\right], \label{dispersion-reduced}
\end{equation}
together with the linear Landau damping rate,  obtained from Eq. \eqref{dispersion-img}, as
\begin{equation}
\gamma_L=-\sqrt{\frac{\pi}{8}}\frac{\omega_pL_q}{\chi^3}\left[1-(q-1)\left(\frac{1}{2\chi^2}+\frac{3}{3q-1}\right)\right]^{(2-q)/(q-1)}, \label{landau-damping-linear-nonextensive}
\end{equation}
where $\omega_p = \sqrt{8\pi n_0 e^2/m}$ is the  plasma oscillation frequency in a charge neutral EP plasma. 
 In the same way, the group velocity expression \eqref{lambda} reduces to 
\begin{equation}
\lambda = \frac{6 \omega_p \chi}{(3q-1)k_d}\left[1+ \frac{45q-11}{(3q-1)(5q-3)}\chi^2\right]. \label{lambda-grop-velocity}
\end{equation}
The dispersion relation \eqref{dispersion-reduced} exactly agrees with that  obtained in Ref. \onlinecite{saberian2013} up to the second order correction term $\propto \chi^2$. However, we observe some interesting features by retaining the term $\propto \chi^4$ in Eq. \eqref{dispersion-reduced} which were overlooked in Ref. \onlinecite{saberian2013}. These interesting  features, as can be seen from Fig. \ref{fig1}, are that the carrier wave dispersion curve and hence the group velocity of the wave envelope can turn over through the $\chi$-axis, going to zero values and then to negative values. These reduction of the wave frequency and the group velocity occurs in the superextensive sub-region $0.47\lesssim q\lesssim3/5$. Beyond this region, i.e., for $q>3/5$, both $\omega_r$ and $\lambda$  increase with increasing values of $\chi$. Furthermore, it is observed that as one goes from the superextensive sub-regime $3/5\lesssim q\lesssim 1$ to subextensive one with $q>1$, the wave frequency and hence the group velocity are seen to get reduced. This is expected since more the superthermal particles the larger is the phase velocity, in agreement with the previous results \cite{saberian2013}.  However, some disagreements are also there for the linear Landau damping rate $\gamma_L$. Here, we  mention that though its analytic expression is the same as obtained in Ref. \onlinecite{saberian2013}, however, the features we observe here are quite distinct. In fact, the possibility of growing instability, as shown in Ref. \onlinecite{saberian2013}, cannot  be made in the superextensive or subextensive plasma regions. From the mathematical expression of $\gamma_L$ one can check that it is always negative regardless of the  values of $q$ and $\chi$ in $1/3<q<1$ (or $q>1$) and $o\lesssim\chi\lesssim1$ respectively. These are  clear from Fig. \ref{fig2}.  Physically, since the phase velocity of the carrier wave is assumed to be larger than the particle's velocity,    the wave modes may lose    energy to the particles instead of gaining energy from them,  and thus be damped.  From Fig. \ref{fig2}, we also find that there are two subregions of $\chi$, namely $0\lesssim\chi\lesssim\chi_0$ and $\chi_0\lesssim\chi\lesssim1$.  In the former,  the damping rate increases, while in the latter  the same decreases with increasing values of $\chi$ (See the left panel). Furthermore, the damping rate is seen to be higher in plasmas with more superthermal particles and  with wavelength in $(0\lesssim\chi\lesssim\chi_0)$. However, the same can be true with higher the number of low-speed particles (or with increasing values of $q$) in the regime  $\chi_0\lesssim\chi\lesssim1$. These are also clear from the right panel of Fig. \ref{fig2}. Here, as $q$ increases, the damping rate increases, however, it gets reduced at long-wavelength perturbations (See the dashed line).


Next, we calculate the various terms (in the limit $\chi^2\ll1$) which appear in the coefficients of the  NLS equation \eqref{nls} for the $q$ distribution \eqref{q-distribution}   as follows:
\begin{equation}
\alpha= -\frac{2k}{\omega_p}\left( 1+ \frac{3}{3q-1}\chi^2\right), \label{alpha-coeff}
\end{equation}
\begin{equation}
\beta=-\frac{6}{k_d(3q-1)}\chi \left[ 1+ \frac{50q-11}{(3q-1)(5q-3)}\chi^2\right], \label{beta-coeff}
\end{equation}
\begin{equation}
A=0,~~W=0,~~U=0, \label{AWU-coeff}
\end{equation}
\begin{equation}
B= -15\left( \frac{e}{m}\right) ^2 \frac{k^3}{{\omega_p}^4}\left( 1+ \frac{24}{3q-1}\chi^2\right),\label{B-coeff}
\end{equation}
\begin{equation}
\Delta= -\frac{2L_qk_d^2}{3\sqrt{\pi(1-q)}}\left( \frac{2q-3}{q-1}\right) \left[ 1-\frac{54(3q^2-5q+3)(q-1)}{(3q-1)^2(2q-3)}\chi^2\right],\label{Delta-coeff}
\end{equation}
\begin{equation}
\Gamma=-\left( \frac{\pi}{2}\right) ^{1/2} k_d^2 \frac{6L_q}{3q-1} \chi\left[ 1- \frac{90q^2-189q+97}{(3q-1)^2(5q-3)}\chi^2\right], \label{Gamma-coeff}
\end{equation}
\begin{equation}
C=\left(  \frac{e}{K_BT}\right) ^2 \frac{1}{k_d^2} \frac{2L_q}{\sqrt{\pi(1-q)}} \frac{4q-5}{3q-3}\left[ 1+\left( \frac{2(216q^2-177q+53)}{(3q-1)^2(4q-5)}-\frac{\sqrt{\pi(1-q)}}{L_q} \frac{45(q-1)}{4q-5}\right) \chi^2\right], \label{C-coeff}
\end{equation}
\begin{equation}
D=-\left(  \frac{e}{K_BT}\right) ^2 \frac{1}{k_d^2} \left( \frac{\pi}{2}\right) ^{1/2} \frac{6L_q}{3q-1} \chi \left[ 1+ \frac{90q^2+21q-61}{(3q-1)^2(5q-3)} \chi^2\right]. \label{D-coeff}
\end{equation}
 Thus, the coefficient $P$ of the NLS equation \eqref{nls} reduces to 
\begin{equation}
P=\frac{3\omega_p}{(3q-1)k_d^2} \left[1+ \frac{3(45q-11)}{(3q-1)(5q-3)} \chi^2\right]. \label{P-reduced}
\end{equation}
To reduce the other coefficients, namely $Q$ and $R$, we first  obtain the expressions for $\Theta(k,\omega)$ and $\Phi(k,\omega)$ in the limit $\chi^2\ll1$. So, we calculate, respectively, the resonant and non-resonant contributions to $\Theta(k,\omega)$ as
\begin{equation}
\frac{1}{\Delta}\frac{k^3}{\Delta^2+\Gamma^2}\left(\Gamma W-\Delta U\right)^2=0, \label{reso-contri}
\end{equation}
\begin{equation}
-k^3\left(\frac{W^2}{\Delta}+C\right)=\left(\frac{e}{k_BT}\right)^2 \frac{2kL_q}{\sqrt{\pi(1-q)}}\frac{4q-5}{3(1-q)}\chi^2. 
\end{equation}
Thus, it turns out that the resonant contribution to the nonlinear coupling coefficient $Q$ being smaller can be disregarded to obtain
\begin{equation}
Q= \frac{1}{2} \left( \frac{e}{k_BT}\right) ^2 \frac{2L_q}{\sqrt{\pi(1-q)}} \frac{4q-5}{3(1-q)}\omega_p\chi^2. \label{Q-reduced}
\end{equation}
However, the contribution from the group velocity resonance through $\Phi(k,\omega)$ gives rise the coefficient $R$ as 
\begin{equation}
R=\frac{3 L_q}{3q-1} \left( \frac{e}{k_BT}\right) ^2 \left( \frac{\pi}{2}\right) ^{1/2} \omega_p\chi^3. \label{R-reduced} 
\end{equation}
Note that the above expressions for $P,~Q$ and $R$ are obtained for $1/3<q<1$. The similar expressions for $q>1$ can also be obtained by replacing the factor $(1-q)$ by $(q-1)$ under the square root in the expressions for $\Delta,~C$ and $Q$. In Sec. \ref{sec-MI}, we will find that though the condition for the modulational instability does not depend on the sign of $PQ$, but on the presence of $R$, however, the sign of $PQ$ are important for determining the values of the frequency shift and the energy transfer rate (in particular, their maximum values) as well as their values in some particular cases, namely when the wave intensity exceeds or smaller than a critical value which  depends on $P,~Q$ and the wave number of modulation. On the other hand,  in Sec. \ref{sec-conservation-laws} we have seen that whether a steady state solution of the NLS equation exists or not depends on the coefficient $R$. Furthermore, in Sec. \ref{sec-soliton-sol} we will also examine the effect of $R$ on solitary wave solution of the NLS equation. Thus, it is useful to investigate the properties of $P,~Q$ and $R$. 

\begin{figure}[ht]
\centering
\includegraphics[height=2.5in,width=7in]{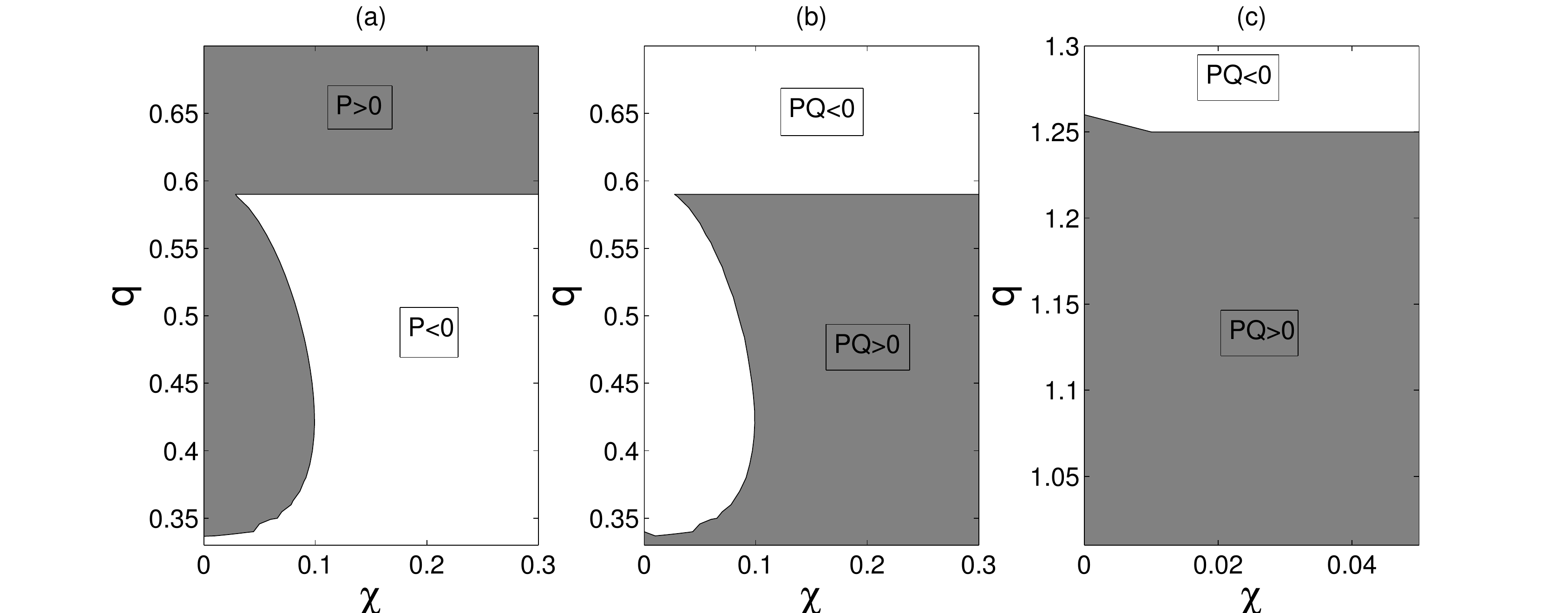}
\caption{(Color online) The regions for $P>0,~PQ>0$ (shaded or gray) and $P<0,~PQ<0$ (blank or white) are shown in the $\chi-q$ $(\chi\equiv k/k_d)$ plane. From the left panel (a), it is seen that $P>0$ for $q\gtrsim3/5$, i.e., in the regimes $3/5\lesssim q\lesssim1$ (superextensive) and $q>1$ (subextensive). However, in the other superextensive regime $1/3<q<3/5$, it can be positive or negative depending on the values of $\chi$ in $0\lesssim\chi\lesssim0.1$. Middle panel (b) shows that in the superextensive regime, $PQ$ is negative for $3/5\lesssim q\lesssim1$, however, it can be positive or negative  in the  sub-regime $1/3<q\lesssim3/5$   depending on the values of $\chi$ in $0\lesssim\chi\lesssim1$. From right panel (c) we find that in subextensive plasmas, $PQ<0$ for $q\gtrsim5/4$ and for $0\lesssim\chi\lesssim1$  except in some small regions of $q~(>5/4)$ and $\chi$ in which $PQ>0$. The latter is also true for $1<q\lesssim5/4$ and $0\lesssim\chi\lesssim1$. Note that $P,~PQ$ are undefined at $q=3/5$.  }
\label{fig3}
\end{figure}

Inspecting on the expressions for $P,~Q$ and $R$, which explicitly depend on the nonextensive parameter $q$ and the nondimensional wave number $\chi~(0\lesssim\chi\lesssim1)$, we find that for superextensive plasmas with $1/3<q\lesssim1$, we have   $Q<0$. However, when $q>1$, i.e., for subextensive plasmas, we have  $Q>0$ $(Q<0)$ in $1<q<5/4$ $(q>5/4)$ and for  $0\lesssim\chi\lesssim1$. Also,   $R>0$ in both the superextensive and subextensive regimes, i.e., $1/3<q\lesssim1$ and $q>1$ with $0\lesssim\chi\lesssim1$. Furthermore, $P$, which has singularity at $q=3/5$, is positive for $q>3/5$, i.e., for $3/5<q\lesssim1$ (superextensive) and $q>1$ (subextensive).  However, in the superextensive subregion $1/3<q<3/5$, $P$ can be negative or positive depending on the rage of values of $\chi$ in $0\lesssim\chi\lesssim1$ (See the left panel of Fig. \ref{fig3}). The regions for $P$ and $PQ$ are shown as contour plots in the $\chi-q$ planes in Fig. \ref{fig3}. 
 We find that in the superextensive regime, $PQ$ is negative for $3/5\lesssim q\lesssim1$, however, it can be positive or negative  in the  sub-regime $1/3<q\lesssim3/5$   depending on the values of $\chi$ in $0\lesssim\chi\lesssim1$ (See the middle panel). From right panel (third from left) we find that in subextensive plasmas, $PQ<0$ for $q\gtrsim5/4$ and for $0\lesssim\chi\lesssim1$  except in some small regions of $q~(>5/4)$ and $\chi$ in which $PQ>0$. The latter is also true for $1<q\lesssim5/4$ and $0\lesssim\chi\lesssim1$. We also note that $P,~PQ$ are undefined at $q=3/5$. 
\section{The nonlinear landau damping and modulational instability} \label{sec-MI}
Here, we follow the same analysis as in Ref. \onlinecite{ichikawa1974}. Though, the relevant analysis is standard, however, we repeat here for the sake of readers.    It is well known that when the group velocity dispersive coefficient $(P)$ and the local nonlinear (cubic) term $(Q)$ of an ordinary NLS equation have the same sign, i.e., $PQ>0$, its  plane wave solution  exhibits instability against a plane wave perturbation of its amplitude and phase \cite{taniuti1969}.  However, the present modified  NLS equation \eqref{nls} contains, in addition to the usual dispersive and nonlinear terms, the nonlocal nonlinear term which is associated with the resonant particles having the group velocity of the wave. Thus, it is  of interest to examine how the nonlocal term contributes to the modulational instability and  describes the nonlinear Landau damping process. Here,  we assume that, before modulation, the NLS equation \eqref{nls} has a plane wave solution of the   form \cite{ichikawa1974,taniuti1969}
\begin{equation}
\phi= \rho^{1/2}\exp\left(i \int ^\xi \frac{\sigma}{2P} d\xi\right), \label{sol-nls}
\end{equation}
where $\rho$ and $\sigma$ are real functions of $\xi$ and $\tau$.  Substituting  the solution \eqref{sol-nls}  into Eq. \eqref{nls} we get  
\begin{equation}
\frac{\partial}{\partial \tau}\rho+ \frac{\partial}{\partial \xi}(\rho\sigma)= -2s\rho, \label{rho}
\end{equation}
\begin{equation}
\frac{\partial}{\partial \tau}\sigma +\sigma\frac{\partial}{\partial \xi} \rho=2PQ\frac{\partial}{\partial \xi}\rho+ \frac{2PR}{\pi}{\cal{P}}\frac{\partial}{\partial \xi} \int \frac{\rho(\xi')}{\xi-\xi'}d\xi' +P^2\frac{\partial}{\partial \xi}\left[ \rho^{-1/2}\frac{\partial}{\partial \xi}\left( \rho^{-1/2}\frac{\partial}{\partial \xi}\rho\right) \right]. \label{sigma}
\end{equation}
As before, we assume that the   linear Landau damping term is higher order than $\epsilon^2$  and linearize   Eqs. \eqref{rho} and  \eqref{sigma} by splitting up $\rho$ and $\sigma$ into their equilibrium (with suffix $0$) and perturbation (with suffix $1$) parts as:
\begin{equation}
\rho= \rho_0 +\rho_1 \cos{(K\xi-\Omega \tau)}+\rho_2\sin{(K\xi-\Omega \tau)}, \label{rho-perturbation}
\end{equation}
\begin{equation}
\sigma= \sigma_1 \cos{(K\xi-\Omega \tau)}+\sigma_2 \sin{(K\xi-\Omega \tau)}, \label{sigma-perturbation}
\end{equation}
where $\Omega$ and $K$ are, respectively, the wave frequency and the  wave number of modulation. 
Now, under  the perturbations \eqref{rho-perturbation} and \eqref{sigma-perturbation}, the solution \eqref{sol-nls} can be expressed as
\begin{eqnarray}
\phi(x,t)=&&\frac{1}{2} \sqrt{\rho_0} \cos{(k_0x-\omega_0t)}\notag\\
&&+A_1(K)\cos{[(k_0+\epsilon K)x-(\omega_0+\epsilon \lambda K+ \epsilon^2 \Omega)t]} \notag\\
&&+A_2(K)\sin{[(k_0+\epsilon K)x-(\omega_0+\epsilon \lambda K+ \epsilon^2 \Omega)t]}\notag \\
&&+ \text{similar terms with}~    K\rightarrow -K~ \text{and}~ \Omega\rightarrow -\Omega, \label{phi-expansion}
\end{eqnarray}
where 
\begin{equation}
A_j(K)= \frac{i}{4\sqrt{\rho_0}}+ \frac{\sqrt{\rho_0}}{4PK}\sigma_j,~~~ j=1,~2.  \label{Aj}
\end{equation}
It follows that due to the linearization of Eqs. \eqref{rho} and  \eqref{sigma}, the electric potential   $\phi$ [Eq. \eqref{phi-expansion}] describes the three-wave interaction of the unperturbed carrier wave $(k_0, \omega_0)$ and two side bands with wave numbers and frequencies $k_0\pm \epsilon|K|$ and $\omega_0\pm \epsilon \lambda|K|\pm \epsilon^2\Omega$.
Now, substituting the perturbation expansions   \eqref{rho-perturbation} and \eqref{sigma-perturbation} into Eqs. \eqref{rho} and \eqref{sigma}, and assuming the smallness of the Landau damping coefficient, i.e., $S=0$, we obtain after eliminating $\sigma_1$ and $\sigma_2$ the following equation
\begin{equation}
\left( \begin{array}{cc}
 \Omega^2 +2\rho_0PQK^2-P^2K^4 & -2\rho_0PR~ \text{sgn} (K)K^2\\
 2\rho_0PR~ \text{sgn} (K)K^2 & \Omega^2 +2\rho_0PQK^2 -P^2K^4\\
\end{array}\right)
\left( \begin{array}{c}
 \rho_1 \\
 \rho_2\\
\end{array}\right) =0, \label{matrix}
\end{equation}
where the coupling between $\rho_1$ and $\rho_2$ appears due to the nonlocal nonlinear term (the second term on the right-hand side) in Eq. \eqref{sigma}. Thus, for nonzero values of $\rho_1$ and $\rho_2$  we obtain the following dispersion relation for electrostatic wave envelopes in collisionless nonextensive EP plasmas. 
\begin{equation}
(\Omega^2 +2\rho_0PQK^2-P^2K^4)^2=-(2\rho_0PR K^2)^2. \label{dispersion-nls}
\end{equation}
The negative sign on the right-hand side of Eq. \eqref{dispersion-nls} shows that whatever be the signs of $P$ and $Q$, the Langmuir wave packet is always unstable due to the nonzero coefficient $R$ associated with the resonant particles having the wave group velocity. Since Eq. \eqref{dispersion-nls} is, in general, complex in $\Omega$,   we seek a general solution of it by considering $\Omega= \Omega_r + i\Gamma $ with $\Omega_r,~\Gamma$ being reals, and obtain
\begin{equation}
\Omega_r=\pm \frac{1}{\sqrt{2}}\left[\left\lbrace \left(P^2 K^2-2\rho_0PQ\right)^2+\left(2\rho_0PR\right)^2\right\rbrace^{1/2} +\left(P^2 K^2-2\rho_0PQ\right)\right]^{1/2}|K|, \label{Omega-real}
\end{equation}
\begin{equation}
\Gamma=\mp\frac{1}{\sqrt{2}}\left[\left\lbrace\left(P^2 K^2-2\rho_0PQ\right)^2+\left(2\rho_0PR\right)^2\right\rbrace^{1/2}-\left(P^2 K^2-2\rho_0PQ\right)\right]^{1/2}|K|, \label{Omega-img}
\end{equation}
where we consider the upper (lower) sign for $K>0~(K<0)$.
 
In what follows, we consider two different limits of the wave amplitude. 
In the  small amplitude limit with $\rho_0\ll |P/2Q|K^2$, Eqs. \eqref{Omega-real} and \eqref{Omega-img} reduce to  
\begin{equation}
\Omega_r\approx\pm PK^2,~\Gamma\approx\mp \rho_0R. \label{Omega-small-amp}
\end{equation}
Since a linear dispersion relation can be expanded as
$\omega(k)=\omega(k_0)+ \omega'(k_0)(k-k_0)+  (1/2)\omega''(k_0)(k-k_0)^2+\cdots$, 
we take the upper sign for $K> 0$ and lower sign for $K< 0$ in Eq. \eqref{Omega-small-amp}. Thus, for $K>0$, we have $\Omega_r=PK^2,~\Gamma=-\rho_0R$ in which the imaginary part $(\propto R)$ is solely due to the resonant  particles having the  group velocity of the wave envelope. It follows that the Langmuir wave packets under the modulation is unstable.    Since $R>0$ for both superextensive $(1/3<q<1)$ and subextensive $(q>1)$ plasmas, this instability is a kind of decay, and of course independent of   $P$ and $Q$. Thus, in the small amplitude limit as above, i.e., when the wave intensity $\rho_0$ is well below a critical value, the real part of $\Omega$ relates the group velocity dispersion, while the imaginary part of $\Omega$ describes   the nonlinear Landau damping process. In the latter,  the wave energy is transferred from the higher frequency side bands to   lower frequency ones. From Eq. \eqref{Omega-small-amp} we also find that     the frequency shift can be positive or negative depending on the values of $q$ and $\chi$ as in Fig. \ref{fig3}.  For example,  $\Omega_r>0$   in the regimes $3/5\lesssim q\lesssim1$ (superextensive) and $q>1$ (subextensive). However, in the other superextensive regime $1/3<q<3/5$, it can be positive or negative depending on the values of $\chi$ in $0\lesssim\chi\lesssim0.1$. Furthermore, for a fixed wave number of modulation $K$, the frequency shift gets significantly reduced in the region $0.47\lesssim q\lesssim3/5$. This is a consequence of the results to the fact that in this region, the carrier wave frequency turn over with the group velocity going to zero values and then to negative values (See Fig. \ref{fig1}). In the other region of $q$, i.e., for $q>3/5$, the frequency $|\omega_r$ increases with $\chi$.  

In the large amplitude limit with $\rho_0\gg |P/2Q|K^2$,  the frequency shift $\Omega_r$ and the transfer rate $\Gamma$ can be obtained as
\begin{equation}
\Omega_r=\pm\sqrt{\rho_0(-PQ)K^2} \left[\sqrt{1+\left({R}/{Q}\right)^2}+1\right]^{1/2}, \label{Omega-real-final}
\end{equation}
\begin{equation}
\Gamma= \mp\sqrt{\rho_0(-PQ)K^2} \left[\sqrt{1+\left({R}/{Q}\right)^2}-1\right]^{1/2}, \label{Omega-img-final}
\end{equation}
which require $PQ<0$. Thus, it turns out that when the wave intensity  $\rho_0$ greatly exceeds a critical value, the frequency shift and the energy transfer rate can be obtained  only in the regions of $PQ<0$ as in the middle and right panels of Fig. \ref{fig3} both in superextensive and subextensive plasmas.  Furthermore,   we note that   $\Omega_r$ and $\Gamma$ are proportional to $\sqrt{\rho_0}$ instead of $\rho_0$ as in the small amplitude case. In particular, for $R=0$ or when the cubic nonlinearity (local) greatly dominates over the local nonlinear term, the modulated wave becomes unstable for $K<K_c\equiv\sqrt{2\rho_0|Q/P|}$ as in the ordinary NLS equation.

Next, from Eq. \eqref{Omega-img} we also find that for a given value of $\rho_0$, the maximum value of the growth rate $\Gamma$ can be achieved at the wave number $K_m$ and  for $PQ>0$ where
\begin{equation}
{K_m}^2= \rho_0\left(\frac{Q^2+R^2}{PQ}\right). \label{K-max}
\end{equation}
The corresponding maximum values of $\Omega_r$ and $\Gamma$ are thus obtained from Eqs. \eqref{Omega-real} and \eqref{Omega-img}  as
\begin{equation}
\Omega_m=\pm \frac{R}{Q}\left(Q^2+R^2\right)^{1/2}\rho_0, \label{Omega-max}
\end{equation}
\begin{equation}
\Gamma_m=\mp\left(Q^2+R^2\right)^{1/2}\rho_0. \label{Gamma-max}
\end{equation}
It follows that the maximum values of the frequency shift and the energy transfer rate for modulated waves can be achieved only in the regions of $q$ and $\chi$ where $PQ>0$. Figure \ref{fig3} (Middle and right panels) confirms that a wide range of values of $\chi$ as well as $q$, both  for superthermal and subextensive plasmas, exist for which $PQ>0$. From the above results we also conclude that in contrast to the ordinary NLS equation (as in fluid theory) in which the modulational instability occurs only for $PQ>0$, the modulated Langmuir wave packets in q-nonextensive plasmas always becomes unstable by the effects of resonant particles having the group velocity of the wave irrespective of the sign of $PQ>0$ or $PQ<0$. In the former, the maximum values of the frequency shift and the growth rate are achieved for arbitrary amplitude of the pump (unperturbed) wave which may not be obtained in Maxwellian plasmas \cite{ichikawa1974}, whereas the latter gives asymptotic values of the same for larger values of the  wave intensity. 

\begin{figure}[ht]
\centering
\includegraphics[height=2.5in,width=5in]{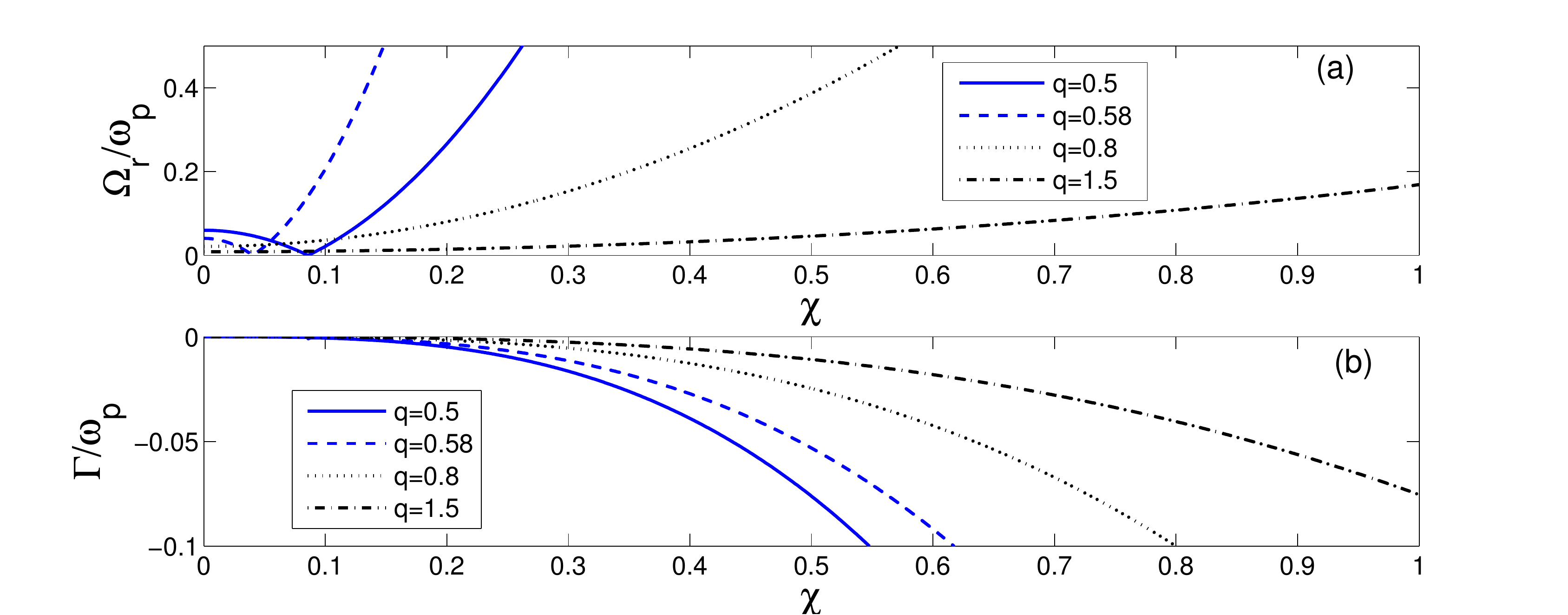}
\caption{ (Color online) The nondimensional frequency shift $\Omega_r/\omega_p$ [upper panel (a)]  and the energy transfer rate $\Gamma/\omega_p$ [lower panel (b)], given by Eqs. \eqref{Omega-real} and \eqref{Omega-img},  are shown with respect to the nondimensional carrier wave number $\chi\equiv k/k_d$   in superextensive $(0.47\lesssim q<1)$ and subextensive $(q>1)$ plasmas. In the superextensive subregion $(0.47\lesssim q<3/5)$ where the  group velocity of the wave envelope vanishes (Fig. \ref{fig1}), the frequency shift and the energy transfer rates are also seen to have cutoffs.   }
\label{fig4}
\end{figure}

In a general manner, we numerically examine the properties of $\Omega_r$ and $\Gamma$ given by Eqs. \eqref{Omega-real} and \eqref{Omega-img} for different values of the nonextensive parameter $q$. The results are displayed in Fig. \ref{fig4} for both  superextensive and subextensive plasmas. From the upper panel of this figure, we find that corresponding to the superextensive regime $0.47\lesssim q\lesssim3/5$ where the carrier wave frequency turn over with the group velocity (Fig. \ref{fig1}), two subregions of $\chi$ exist, in one of which    $\Omega_r$ decreases having cutoffs at lower  $\chi$, while it increases with increasing values of $q$ and $\chi$. In the other  region of $q$, namely, $q>3/5$, the frequency shift is seen to increase with $\chi$ without any cutoff, while it decreases with increasing values of $q$. For the damping rate $\Gamma$, some different features are observed (See the lower panel). Here, as $q$ increases, the values of $|\Gamma|$ decrease, i.e., more the number of superthermal particles the higher is the rate of energy transfer from high-frequency side bands to low-frequency ones.

\section{Nonlinear Landau damping of solitary wave solution}\label{sec-soliton-sol}
In absence of both the linear and nonlinear Landau damping effects, i.e., for $R=S=0$,  the modified NLS equation \eqref{nls} reduces to the following ordinary NLS equation 
\begin{equation}
i\frac{\partial\phi}{\partial \tau}+P \frac{\partial^2\phi}{\partial \xi^2}+Q |\phi|^2 \phi=0. \label{nls-usual}
\end{equation}
In the case   where the modulational instability occurs for $PQ>0$, a stationary solution (bright soliton) of Eq. \eqref{nls-usual} can be obtained as \cite{fedele2002}
\begin{equation}
\phi=\phi_0 \exp{(i\theta)}, \label{soliton-ordinary-nls}
\end{equation}
where the amplitude $\phi_0$ and the phase $\theta$ are given by
\begin{equation}
\phi_0=\sqrt{\widetilde{\phi_0}}~\text{sech} \left( \frac{\xi-v_0\tau}{L}\right), \label{phi-0}
\end{equation}
\begin{equation}
\theta=\frac{1}{2P}\left[v_0\xi+ \left(\Omega- \frac{v_0^2}{2}\right)\tau\right], \label{theta}
\end{equation}
 with $L\widetilde{\phi_0}=\sqrt{2|P/Q|}$ being a constant and $z=(\xi-v_0\tau)/L$ denoting the transformation in the moving (with velocity $v_0$) frame of reference.  
 
 In this section our aim is to determine the effect of a small amount of the nonlinear Landau damping $(R)$ associated with the resonant particles having the group velocity of the wave envelope on the soliton  solution \eqref{soliton-ordinary-nls}. Here, we  disregard the small effect of the linear Landau damping rate $\propto S$ which has been assumed to be  higher order than $\epsilon^2$. From Eq. \eqref{dI2-dtau} it is evident that an initial perturbation of the form \eqref{soliton-ordinary-nls} will decay to zero. Thus, one might expect that  the amplitude $\widetilde{\phi_0}$ is no longer a constant but can decrease slowly  with time, i.e.,  $\widetilde{\phi_0}=\widetilde{\phi_0}(z,\tau)$.
  We consider $|P|,~|Q|\gg|R|\sim\epsilon\gg S\sim\epsilon^{2+p},~p>0$ and do the perturbation analysis of the NLS equation \eqref{nls} with $R$ as the small parameter. It can be easily  verified that this assumption is  valid for both superextensive and subextensive regions of $q$ as mentioned before. We follow the similar approach as has been applied  in different studies, however, to Korteweg-de Vries (KdV) equations (See, e.g., Refs. \onlinecite{misra2014,misra2015} and references therein).
Now, under the   transformation $z=(\xi-v_0\tau)/L$, Eq. \eqref{nls} reduces to 
\begin{equation}
i\frac{\partial\phi}{\partial \tau}-i\frac{v_0}{L}\frac{\partial \phi}{\partial z}+\frac{P}{L^2} \frac{\partial^2 \phi}{\partial z^2}+Q |\phi|^2 \phi +\frac{r}{\pi}{\cal{P}}\int\frac{ |\phi(z')|^2}{z-z'} dz'\phi=0, \label{nls1}
\end{equation}
where we rewrite  $R\equiv r$   to denote $R$ as small (We will replace again $r$ by $R$ in the final solution) and  $\partial\phi/\partial z'=\partial\phi/\partial z$ at $z=z'$.

In what follows, to  investigate the solution of Eq. \eqref{nls1} we generalize the multiple time scale analysis with respect to $r$, i.e.,    we consider the solution as
\begin{equation}
\phi(z,\tau)= \phi^{(0)}+ r \phi^{(1)}+r^2 \phi^{(2)}+\cdots \label{sol-phi-z-tau}
\end{equation}
where $\phi^{(i)},~ i=0,1,2,3,...$, are functions of $\tau= \tau_0, \tau_1, \tau_2,...$
Substituting  Eq. \eqref{sol-phi-z-tau} into Eq. \eqref{nls1} and equating the coefficients of the zeroth and first orders of $r$, we successively obtain
\begin{equation}
i\frac{\partial \phi^{(0)}}{\partial \tau}-i\frac{v_0}{L}\frac{\partial \phi^{(0)}}{\partial z}+\frac{P}{L^2}\frac{\partial^2 \phi^{(0)}}{\partial z^2}+Q|\phi^{(0)}|^2\phi^{(0)}=0, \label{r0}
\end{equation}
\begin{equation}
i\frac{\partial \phi^{(1)}}{\partial \tau}+\left(\frac{\partial}{\partial z}\Lambda_1+Q|\phi^{(0)}|^2\right)\phi^{(1)}+Q\left(\phi^{(0)}\right)^2\phi^{(1)\ast}=\Lambda_2\phi^{(0)}, \label{r1}
\end{equation}
where $\Lambda_1$ and $\Lambda_2$ are given by
\begin{equation}
\Lambda_1=\frac{P}{L^2}\frac{\partial}{\partial z}-i\frac{v_0}{L},~~ 
\Lambda_2=-\left(i\frac{\partial}{\partial \tau_1}+\frac{1}{\pi}{\cal{P}}\int\frac{|\phi^{(0)}(z')|^2}{z-z'} dz'\right). 
\end{equation}
\par
It can   easily be shown that under the boundary conditions, namely $\phi^{(0)},~\partial \phi^{(0)}/\partial z,~\partial^2 \phi^{(0)}/\partial z^2\rightarrow0$ as $z\rightarrow\infty$, Eq. \eqref{r0} possesses a solution of the form  $ \phi^{(0)}=\sqrt{\widetilde{\phi}_0}~ { \text{sech}}~{z} \exp{\left(i\theta_1z\right)}$  $\iff$    ${\partial \phi^{(0)}}/{\partial \tau}=0$ for some real values of $\theta_1$. 
Now, for the existence of   solution of Eq. \eqref{r1}, it is necessary that $\Delta\phi^{(0)}$ be orthogonal to all solution $g(z)$ of $L^+[g]=0$ which satisfy $g(\pm \infty)=0$, where $L^+$ is the operator adjoint to $L$ defined by
\begin{equation}
\int_{-\infty}^{\infty} \psi_1(z)L[\psi_2(z)]dz=\int_{-\infty}^{\infty} \psi_2(z)L^+[\psi_1(z)]dz, \label{177 eq}
\end{equation}
with $\psi_1(\pm \infty)=\psi_2(\pm \infty)=0$, and the only solution of $L^+[g]=0$ is $g(z)=\text{sech}~{ z}$.
Thus,  we have
\begin{equation}
\int_{-\infty}^{\infty}\Lambda_2\phi^{(0)}\text{sech}~{z}dz=0,
\end{equation}
which gives
\begin{equation}
\frac{i}{2\sqrt{\widetilde{\phi}_0}} \frac{\partial \widetilde{\phi}_0}{\partial \tau_1} \left[ \frac{\sqrt{2\pi}\theta_1 \sinh {\left({\pi \theta_1}/{2}\right)}}{1-\cosh{(\pi \theta_1)}}\right]-\frac{1}{\pi}\widetilde{\phi}_0^{3/2}{\cal P}\int_{-\infty}^{\infty}\int_{-\infty}^{\infty}\left(\frac{\text{sech}^2 z'}{z-z'}\right) \text{sech}^2z \exp(i\theta_1z)dz dz'=0. \label{eq1-phi0-tilde}
\end{equation}
Equation \eqref{eq1-phi0-tilde} is a first-order differential equation for the wave amplitude $\widetilde{\phi}_0$. So, its solution can be obtained as 
\begin{equation}
\widetilde{\phi}_0=\widetilde{\phi}_{00} \left(1-i\frac{\tau}{\tau_0}\right)^{-1},
\end{equation}
where $\widetilde{\phi}_0=\widetilde{\phi}_{00}$ at $\tau=0$ and  $\tau_0$ is given by (rewriting now $r$ as $R$)
\begin{equation}
\tau_0^{-1}=\frac{\sqrt{2}R\widetilde{\phi}_{00}}{{\pi}^{3/2}\theta_1}\left[\frac{\cosh{(\pi\theta_1)}-1}{ \sinh{\left({\pi\theta_1}/{2}\right)}}\right]  {\cal{P}}  \int_{-\infty}^{\infty}\int_{-\infty}^{\infty}\left(\frac{\text{sech}^2 z'}{z-z'}\right) \text{sech}^2z \exp(i\theta_1z)dz dz'.  
\end{equation}
Thus, when $PQ>0$, an approximate solitary wave solution of the NLS equation \eqref{nls} with a small effect of the  nonlinear Landau damping is given by
\begin{equation}
\phi=\sqrt{\widetilde{\phi}_{00}}\left(1-i\frac{\tau}{\tau_0}\right)^{-1/2}\text{sech}~{z}\exp(i\theta), \label{sol1-approx-nls}
\end{equation}
where $\theta= \left[v\xi+\left(\Omega-{v_0^2}/{2}\right)\tau\right]/2P$.

On the other hand, when $PQ<0$,  a stationary solution (dark soliton) of Eq. \eqref{nls-usual} can be taken as \cite{fedele2002}
 \begin{equation}
\phi=\phi_0 \exp(i\theta),
\end{equation}
where $\phi_0$ and $\theta$ are different from those given by Eqs. \eqref{phi-0} and \eqref{theta}, i.e.,  
\begin{equation}
\phi_0=\widetilde{\phi}_0\tanh\left( \frac{\xi-v_0\tau}{L}\right),
\end{equation}
\begin{equation}
\theta=\frac{1}{2P}\left[v_0\xi+ \left(2PQ\widetilde{\phi}_0^2-\frac{v_0^2}{2}\right)\tau\right], \label{theta-PQ<0}
\end{equation} 
with $L\widetilde{\phi}_0=\sqrt{|{2P}/{Q}|}=$ a constant.

Proceeding in the same way as above up to Eq. \eqref{r1}, it can be easily verified that 
$\phi^{(0)}=\widetilde{\phi}_0\tanh{z} \exp{(i\theta_2 z)}$, for some $\theta_2$, is a solution of Eq. \eqref{r0} if and only if ${\partial \phi^{(0)}}/{\partial \tau}=0$. Also, for the existence of a solution of 
Eq. \eqref{r1} we have the same necessary condition but with different $g(z)=\tanh{z}$. The condition gives 
\begin{equation}
\int_{-\infty}^{\infty}\Lambda_2\phi^{(0)}\tanh{z}dz=0,
\end{equation}
from which we obtain
\begin{equation}
 i \sqrt{2\pi}\frac{\partial \tilde{\phi_0}}{\partial \tau_1} \left[\delta(t)+ \frac{\theta_2 \sinh \left({\pi \theta_2}/{2}\right)}{1-\cosh(\pi \theta_2)}\right] +\frac{1}{\pi}\widetilde{\phi}_0^3{\cal{P}}\int_{-\infty}^{\infty}\int_{-\infty}^{\infty}\left(\frac{\tanh^2 z'}{z-z'}\right) \tanh^2 z \exp(i\theta_2z)dz dz'=0, \label{eq2-phi0-tilde}
\end{equation}
where $\widetilde{\phi}_0=\widetilde{\phi}_{00}$ at $\tau=0$. As before, the solution of Eq. \eqref{eq2-phi0-tilde} in the case of $PQ<0$ is given by 
\begin{equation}
\widetilde{\phi}_0=\widetilde{\phi}_{00}\left(1-i\frac{\tau}{\tau_0}\right)^{-1/2},
\end{equation}
where (Rewriting $r$ as $R$)
\begin{equation}
\tau_0^{-1}=\left(\frac{2}{\pi}\right)^{3/2}\frac{R\widetilde{\phi}_{00}^2\left[1-\cosh(\pi\theta_2)\right]}{\delta(\tau)(1-\cosh\pi\theta_2)+\theta_2 \sinh\left(\frac{\pi\theta_2}{2}\right)}  {\cal{P}} \int_{-\infty}^{\infty}\int_{-\infty}^{\infty}\left(\frac{\tanh^2 z'}{z-z'}\right) \tanh^2z \exp(i\theta_2z)dz dz',
\end{equation}
with $\delta(\tau)$ denoting the Dirac delta function.
Thus,  when $PQ<0$ an approximate solitary wave solution of the NLS equation \eqref{nls} with a small effect  of the nonlinear Landau damping is given  by
\begin{equation}
\phi={\widetilde{\phi}_{00}}\left(1-i\frac{\tau}{\tau_0}\right)^{-1/2}\text{tanh}{z}\exp(i\theta), \label{sol2-approx-nls}
\end{equation}
where $\theta$ is given by Eq. \eqref{theta-PQ<0}.

\begin{figure}[ht]
\centering
\includegraphics[height=2.5in,width=5in]{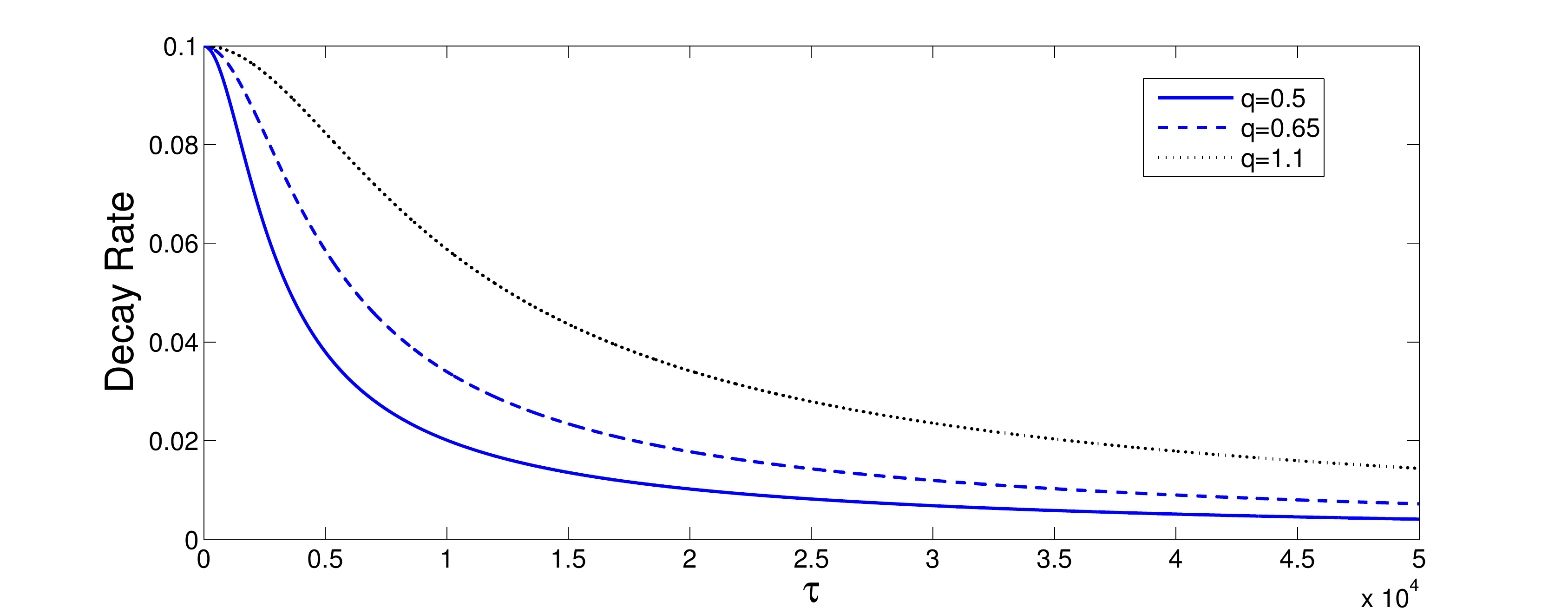}
\caption{(Color online) The nondimensional decay rate $|\widetilde{\phi}_0/\widetilde{\phi}_{00}|$ is shown  for both superextensive (solid and dashed lines) and subextensive (dotted line) plasmas. It is seen that faster the decay of the wave amplitude, the larger is the percentage of superthermal particles in plasmas. }
\label{fig5}
\end{figure}

From Eqs. \eqref{sol1-approx-nls} and \eqref{sol2-approx-nls}, it is evident that the  absolute value  $(|\phi|)$ of the wave amplitude decays slowly with time $\tau$ with a small effect of the nonlinear Landau damping.  Figure \ref{fig5} exhibits a qualitative plot of the absolute value of $\phi$ given by Eq. \eqref{sol1-approx-nls} in plasmas with superextensive and subextensive velocity distributions. It is seen that faster the decay of the wave amplitude, the larger is the percentage of superthermal particles in plasmas. 
\section{Conclusion} We have investigated the amplitude modulation and the nonlinear evolution of electrostatic wave envelopes in a collisionless   electron-positron pair plasma in the context of Tsallis' $q$ nonextensive statistics. Starting from a set of Vlasov-Poisson equations and applying the reductive perturbation technique, the dynamics of the wave envelopes is shown to be governed by a NLS equation with a nonlocal nonlinear term arising from resonant particles having the group velocity of the wave envelope. Such  wave-particle resonance   also   modifies the local nonlinear (cubic) coupling coefficient of the NLS equation. Furthermore, the nonextensive parameter $q$, which measures the excess of superthermal particles in plasmas, is   shown to modify the dispersive (group velocity), local nonlinear (cubic) as well as the nonlocal nonlinear terms significantly. An interesting effect of $q$ is that a subregion $(0.47\lesssim q\lesssim3/5)$ of the superextensivity $(1/3<q<1)$ exists where the carrier wave dispersion curve $(\omega_r)$ can turn over with the group velocity $(\lambda)$ going to zero and then to  negative values (Fig. \ref{fig1}). Such features of the dispersion curve have not been reported in the previous work  \cite{saberian2013} where the same plasma model has been considered to investigate linear Landau damping of Langmuir oscillations in $q$-nonextensive EP plasmas. In Ref. \onlinecite{saberian2013} these effects were absent due to   truncation of the wave frequency   up to $\chi^2$. Furthermore, our results in the linear theory  show that the electrostatic wave is always damped due to   resonant particles having phase velocity of the wave in both superextensive and subextensive regimes and also for long-wavelength perturbations. These are also in disagreement with the results of Ref. \onlinecite{saberian2013} where the possibility of growing instability has been predicted. We, however, stress that such a growing instability should not appear for high-frequency oscillations in $q$-nonextensive EP pair plasmas. This is due to the fact that since the wave phase velocity is assumed to be larger than the particle's velocity,  the wave modes can be damped by losing their energy to the particles.      It is found that for a fixed $q$, two subregions of $\chi$ exist, in one of which the linear damping rate $\gamma_L$ becomes higher, while in the other it gets reduced with increasing values of $\chi$. On the other hand, for a fixed $\chi$, the absolute value of $\gamma_L$ increases with increasing values of $q$, however, its value gets reduced in larger wavelength of perturbations.

  In the nonlinear regime, we have verified the conservation laws as applicable for a NLS equation. It is found that unlike the ordinary NLS equation, the nonlocal nonlinear term, associated with the nonlinear wave-particle resonance, violates the conservation laws, leading  to a decay of the wave amplitude with time and thereby  forbidding the existence of a steady state solution of the modified NLS equation \eqref{nls}.   We show that the modulated wave packet is always unstable  (regardless of the sign of $P$ and $Q$) due to the nonlocal nonlinear term which can describe the nonlinear Landau damping process in which the wave energy is transferred from   higher to lower frequency sidebands.
      
  The frequency shift $(\Omega_r)$ and the energy transfer rate $(\Gamma)$ for the modulated waves are also examined by the parameter $q$ in a general way as well as in the limits of small $(\rho_0\ll\rho_c)$ and large $(\rho_0\gg\rho_c)$  amplitudes, where $\rho_c=|P/2Q|K^2$ is some critical value of the pump wave intensity $\rho_0$. It is found that both $\Omega_r$ and $\Gamma$ attain their maximum values only when $PQ>0$.   Such maxima of $\Omega_r$ and  $\Gamma$ may not exist in pair plasmas or electron-ion plasmas with Maxwellian distributions  \cite{ichikawa1974}.  However, they assume some asymptotic values  in the limit $\rho_0\gg\rho_c$ and when $PQ<0$, which are $\propto\sqrt{\rho_0}$ instead of $\rho_0$ as in the opposite limit $\rho_0\ll\rho_c$. The regions of $q$ and $\chi$ for which $P,~PQ>0$ and $P,~PQ<0$ are also obtained in both superextensive $(q<1)$ and subextensive $(q>1)$ regimes. 
  
 The general expressions of both $\Omega_r$ and  $\Gamma$  are also studied   by the effects of  $q$.  It is found that corresponding to the superextensive regime $0.47\lesssim q\lesssim3/5$ where the carrier wave frequency vanishes with the group velocity (Fig. \ref{fig1}), two subregions of $\chi$ exist, in one of which a significant reduction of   $\Omega_r$  occurs having cutoffs at lower  $\chi$, while it gets enhanced with increasing values of $q$ and $\chi$. The existence of such cutoffs are the consequences of the turn over effects  of the carrier wave frequency as well as the group velocity of the wave envelope. In the other  region of $q$, namely, $q>3/5$, the frequency shift is seen to increase with $\chi$ (having no cutoff), while it decreases with increasing values of $q$.  However, quite distinct   features are observed  for the energy transfer rate $\Gamma$. Here, as $q~(>1/3)$ increases, the values of $|\Gamma|$ decrease, i.e., more the percentage of superthermal particles the higher is the rate of energy transfer from high-frequency side bands to low-frequency ones.
 
  We have also studied the effects of a small amount of the nonlinear Landau damping $(\propto R)$ on the solitary wave solution of  the ordinary NLS equation by assuming that the linear Landau damping $\propto \gamma_L$ rate is higher order than $\epsilon^2$.    It is found that the wave amplitude decays with time and the  decay rate  can be faster the larger is the number of superthermal particles in EP  plasmas. To conclude, the results should be useful for the evolution of nonlinear wave envelopes and associated wave damping in collisionless pure EP plasmas or pure pair-ion plasmas such those in laboratory \cite{liu1994}, space \cite{montgomery1968} and astrophysical \cite{pierrard2010} environments.
\acknowledgments{This work was supported by UGC-SAP (DRS, Phase III) with  Sanction  order
No.  F.510/3/DRS-III/2015(SAPI) dated  25/03/2015, and UGC-MRP with F. No. 43-539/2014 (SR) and FD Diary No. 3668 dated 17.09.2015.}
\appendix
\section{Modes with $n=l=1$}\label{appendix1}
Equating the coefficients of $\epsilon$ from Eqs. \eqref{vlasov1} and \eqref{poisson1} for $n=1,~l=1$  we successively obtain 
 \begin{equation}
 f^{(1)}_{\alpha,1}=-k\phi^{(1)}_1\frac{G_\alpha (v)}{\omega-kv},~k^2\phi^{(1)}_1=4\pi\sum e_\alpha\int_C f^{(1)}_{\alpha,1}dv. \label{f-phi-1}
 \end{equation} 
 For nonzero perturbations, Eq. \eqref{f-phi-1}  yields the following linear dispersion law:
 \begin{equation}
 k+4\pi\sum e_\alpha\int_C\frac{G_\alpha(v)}{\omega-kv}dv=0. \label{dispersion-reln}
\end{equation}
 The small parameter $\epsilon$, which measures the amplitude of the wave, can be  related to the linear Landau damping rate $\gamma_L$  of the plasma wave as  $\gamma_L\sim\omega_r\epsilon^{2+p}$, where $p$ is a nonnegative integer and $\omega_r$ is the real part of $\omega= \omega_r+i\gamma_L$. 
Next, from Eq. \eqref{dispersion} we have the plasma dielectric function
\begin{equation}
D(k, \omega)\equiv1-\frac{4\pi e^2}{mk^2}\int \frac{ \frac{\partial}{\partial v}\left[F^{(0)}_e (v)+F^{(0)}_p (v)\right]}{v-\omega/k}=0,\label{Dk-omega}
\end{equation}
where for pair plasmas or EP plasmas $m_e=m_p=m$.  If $\gamma_L\ll\omega_r$, i.e.,  $\gamma_L\sim\omega_r\epsilon^{2+p}$,  then the dielectric function $D(k, \omega)$ can  be Taylor expanded in the smallness of $\gamma_L$ to obtain its real (with suffix r) and imaginary (with suffix i) parts as \cite{saberian2013} 
\begin{equation}
D_r(k, \omega_r)=1-\frac{4\pi e^2}{mk^2} \int_C \frac{ \frac{\partial}{\partial v}\left[F^{(0)}_e (v)+F^{(0)}_p (v)\right]}{v-\omega_r/k}, \label{dielectric real fnc}
\end{equation}
\begin{equation}
D_i(k,\omega_r)=-\pi\left( \frac{4\pi e^2}{mk^2}\right) \left[ \frac{\partial}{\partial v}\left\lbrace F^{(0)}_e (v)+F^{(0)}_p (v)\right\rbrace\right]_{v=\omega_r/k}. \label{dielectric im fnc}
\end{equation}
Next,  neglecting the terms of order $\left(\gamma_L/\omega_r\right)^2$ and higher in Eqs. \eqref{dielectric real fnc} and \eqref{dielectric im fnc}, the expressions for $\omega_r$ and $\gamma_L$ can be obtained   from the following relations 
 \begin{equation}
 D_r(k, \omega_r)=0, \label{dispersion-real}
 \end{equation}
 \begin{equation}
 \gamma_L=- \frac{D_i(k, \omega_r)}{\partial D_r(k, \omega_r)/ \partial \omega_r}\equiv\frac{\pi}{k}\sum_{\alpha}e_{\alpha}G_{\alpha}\left(\frac{\omega}{k}\right)/\sum_{\alpha}e_{\alpha}\int_C\frac{G_{\alpha}}{\left(\omega-kv\right)^2}dv. \label{dispersion-imagi}
 \end{equation}
\section{Modes with $n=1,~l\neq0$}\label{appendix2}
We equate the components for $l\neq 0$ and $n=1$. Thus, from Eq. \eqref{vlasov1} we have
\begin{equation}
il(\omega-kv)f^{(1)}_{\alpha,l}+ilkG_\alpha \phi^{(1)}_l=0.
\end{equation}
This gives
\begin{equation}
 f^{(1)}_{\alpha,l}\doteq-\frac{G_\alpha}{\omega-kv+i\nu}k\phi^{(1)}_l, \label{f-alpha-l1}
\end{equation}
where $\nu=\pm |\nu|$ for $l\lessgtr0$ has been introduced to  anticipate that the solution in the linear approximation decays with Landau damping rate. 

Similarly, for $l\neq 0$ and $n=1$   equating the coefficients of $\epsilon$ from Eq. \eqref{poisson1} we obtain 
\begin{equation}
l^2k^2 \phi^{(1)}_l-4\pi\sum e_\alpha \int f^{(1)}_{\alpha,l}dv=0. \label{phi1}
\end{equation}
Next, substituting Eq. \eqref{f-alpha-l1} into Eq. \eqref{phi1} and noting that
 \begin{equation}
 \int_R dv\sim \int_C dv\pm i\pi \int dv \delta(\omega-kv) \sim\int_C dv\pm o(\epsilon^{2+p}),
 \end{equation}
we obtain 
  \begin{equation}
 \left( l^2k+4\pi\sum e_\alpha \int_C \frac{G_\alpha}{\omega-kv}dv\right)  \phi^{(1)}_l=0,  \label{phi-l-1}
 \end{equation}
 where the contour $C$ enables us to remove $\nu$ in the denominator of Eq. \eqref{phi-l-1}. Thus, by means of the dispersion relation \eqref{dispersion}, one must have
 \begin{equation}
 f^{(1)}_{\alpha,l}\doteq 0~\text{and}~\phi^{(1)}_l=0~\text{for}~|l|\geq 2. \label{condition-f-phi-l-1}
 \end{equation}

\section{Expressions for $n=2$ and $l=1$}\label{appendix3}
We consider   the second order, first harmonic modes with $n=2$ and $l=1$. Thus,   from Eq. \eqref{vlasov1} we have
\begin{equation}
i(\omega - kv)f^{(2)}_{\alpha,1} +ikG_\alpha \phi^{(2)}_1 = \frac{\partial}{\partial \sigma}
f^{(1)}_{\alpha,1} +v \frac{\partial}{\partial \eta} f^{(1)}_{\alpha,1}-G_\alpha\frac{\partial}{\partial \eta} \phi^{(1)}_1,
 \end{equation} which gives
\begin{equation}
 f^{(2)}_{\alpha,1}\doteq -\frac{1}{\omega -kv+i\nu} \left[kG_\alpha \phi^{(2)}_1+i\left(  \frac{\partial}{\partial \sigma}
f^{(1)}_{\alpha,1} +v \frac{\partial}{\partial \eta} f^{(1)}_{\alpha,1}-G_\alpha\frac{\partial}{\partial \eta} \phi^{(1)}_1 \right) \right]. \label{f-vlasov-n2-l1}
\end{equation}
Also,   from Eq. \eqref{poisson1} we have for $n=2$ and $l=1$ the following equation
\begin{equation}
k^2\phi^{(2)}_1-2ik \frac{\partial}{\partial \eta} \phi^{(1)}_1-4\pi\sum e_\alpha \int_C f^{(2)}_{\alpha,1}dv=0. \label{phi-poisson-n2-l1} 
\end{equation}
This equation can be reduced, by substituting from Eq. \eqref{f-vlasov-n2-l1} the expression of $f^{(2)}_{\alpha,1}$ into it and using the dispersion relation \eqref{dispersion-reln} and the expression \eqref{f-alpha-l1}, to
 \begin{equation}
\frac{\partial}{\partial \eta} \phi^{(1)}_1 +4\pi\sum e_\alpha \int_C\frac{G_\alpha}{(\omega-kv)^2} \left[\frac{\partial}{\partial \sigma}\phi^{(1)}_1+v \frac{\partial}{\partial \eta}\phi^{(1)}_1\right] dv=0, \label{phi-poisson-n2-l1-reduced}
\end{equation}
Equation \eqref{phi-poisson-n2-l1-reduced}  can be written in the form
 \begin{equation}
\left\{ \frac{\partial}{\partial \sigma} + \lambda \frac{\partial}{\partial \eta}\right\} \phi^{(1)}_1(\eta, \sigma;\zeta) =0. \label{vg_eq}
\end{equation}
\section{Second order harmonic modes for $n=l=2$}\label{appendix4}
For   $n=l=2$ we have  from  Eqs. \eqref{vlasov1}
\begin{equation}
2i(\omega-kv)f^{(2)}_{\alpha,2}+2ik G_\alpha\phi^{(2)}_2\doteq \frac{\partial}{\partial \sigma}f^{(1)}_{\alpha,2} +v\frac{\partial}{\partial \eta}f^{(1)}_{\alpha,2}-G_\alpha \frac{\partial}{\partial \eta} \phi^{(1)}_2 -ik \frac{e_\alpha}{m_\alpha}\phi^{(1)}_1\frac{\partial}{\partial v}f^{(1)}_{\alpha,1},
\end{equation}  which by means of Eqs. \eqref{f-alpha-l1} and \eqref{condition-f-phi-l-1} yields
\begin{equation}
 f^{(2)}_{\alpha,2}\doteq - \frac{k}{\omega-kv+i\nu}\left[G_\alpha \phi^{(2)}_2 -\frac{e_\alpha}{2 m_\alpha}k \frac{\partial}{\partial v}\left(\frac{G_\alpha}{\omega-kv+i\nu}\right) \left(\phi^{(1)}_1\right)^2\right]. \label{f_alpha2-n2-l2}
\end{equation}
Next, the expression for $\phi^{(2)}_2$ is obtained  from Eq. \eqref{poisson1} for $n=l=2$ and using the relation \eqref{f_alpha2-n2-l2} as
\begin{equation}
4k^2 \phi^{(2)}_2-4ik\frac{\partial}{\partial \eta}\phi^{(1)}_2 -4\pi\sum e_\alpha \int f^{(2)}_{\alpha,2}dv=0,
\end{equation}
i.e., 
\begin{equation}
\phi^{(2)}_2=\frac{1}{6} A(k,\omega)\left(\phi^{(1)}_1\right)^2, \label{phi2_n2-l2}
\end{equation}
\section{Harmonic modes with $n=3,~l=0$}\label{appendix5}
For  $n=3$ and $l=0$ we have from Eq. \eqref{vlasov1}
\begin{eqnarray}
&&\frac{\partial}{\partial \sigma}f^{(2)}_{\alpha,0}+v\frac{\partial}{\partial \eta}f^{(2)}_{\alpha,0}-G_\alpha \frac{\partial}{\partial \eta} \phi^{(2)}_0=-ik\frac{e_\alpha}{m_\alpha}\phi^{(2)}_{-1}\frac{\partial}{\partial v}f^{(1)}_{\alpha,1}+ik\frac{e_\alpha}{m_\alpha}\phi^{(2)}_1\frac{\partial}{\partial v}f^{(1)}_{\alpha,-1} \notag\\ &&-ik\frac{e_\alpha}{m_\alpha}\phi^{(1)}_{-1}\frac{\partial}{\partial v}f^{(2)}_{\alpha,1} +ik\frac{e_\alpha}{m_\alpha}\phi^{(1)}_1\frac{\partial}{\partial v}f^{(2)}_{\alpha,-1}
 +\frac{e_\alpha}{m_\alpha}\frac{\partial}{\partial \eta}\phi^{(1)}_{-1}\frac{\partial}{\partial v}f^{(1)}_{\alpha,1}+\frac{e_\alpha}{m_\alpha}\frac{\partial}{\partial \eta}\phi^{(1)}_1\frac{\partial}{\partial v}f^{(1)}_{\alpha,-1}, 
\end{eqnarray}
where we have used the relations \eqref{f-phi-n1-l0}. This equation can further be reduced after few steps using Eqs. \eqref{f-alpha-l1}, \eqref{f-vlasov-n2-l1} and \eqref{vg_eq} to the following equation: 
\begin{equation}
\frac{\partial}{\partial \sigma}f^{(2)}_{\alpha,0}+v\frac{\partial}{\partial \eta}f^{(2)}_{\alpha,0}-G_\alpha \frac{\partial}{\partial \eta} \phi^{(2)}_0\doteq \frac{e_\alpha}{m_\alpha}k^2 I_\alpha(v) \frac{\partial}{\partial \eta}|\phi^{(1)}_1|^2, \label{f-alpha-n2-l0}
\end{equation}
where
\begin{equation}
 I_\alpha(v)=\frac{\partial}{\partial v}\left\{\frac{v-\lambda}{(\omega-kv)^2}G_\alpha\right\}. \label{I-alpha-v}
\end{equation}
Next, taking the Fourier-Laplace transform of Eq. \eqref{f-alpha-n2-l0} with respect to $\eta$ and $\sigma$ and using the  initial condition  \eqref{int-cond1} we obtain
\begin{equation}
\tilde{f}^{(2)}_{\alpha,0}(v,K,\Omega,\zeta)\doteq -\frac{K}{\Omega-Kv}G_\alpha(v) \tilde{\phi}^{(2)}_0(K,\Omega,\zeta) -k^2\frac{e_\alpha}{m_\alpha}\frac{K}{\Omega-Kv}I_\alpha(v)H(K,\Omega), \label{fF-alpha-n2-l0}
\end{equation}
where $H(K,\Omega)$ is defined as
\begin{equation}
 \mid\phi^{(1)}_1(\eta-\lambda \sigma,\zeta) \mid^2=\frac{1}{(2\pi)^2}\int d\Omega \int dKH(K,\Omega)\exp[i(K\eta-\Omega\sigma)], \label{phi-1-1-mod}
\end{equation}
with
\begin{equation}
H(K,\Omega)=2\pi\delta(\Omega-K\lambda)\int dK' \phi^{(1)\ast}_1(K')\phi^{(1)}_1(K+K'). \label{H-K-Omega}
\end{equation}
Now, for $n=2,~l=0$ Eq. \eqref{poisson1} reduces to 
\begin{equation}
4\pi \sum e_\alpha\int \tilde{f}^{(2)}_{\alpha,0}=0.  
\end{equation}
Thus,  we obtain from Eq. \eqref{fF-alpha-n2-l0} the following relation
\begin{equation}
\tilde{\phi}^{(2)}_0=k^2\frac{H(K,\Omega)}{\Delta^{(c)}(K,\Omega)}{\cal{W}} (K,\Omega), \label{phiF-n2-l0}
\end{equation}
where 
\begin{equation}
{\cal{W}}(K,\Omega)=4\pi \sum_\alpha \frac{e^2_\alpha}{m_\alpha} \int \frac{K}{(\omega-kv)^2(\Omega-Kv)}G_\alpha dv, \label{W-cal}
\end{equation}
\begin{equation}
\Delta^{(c)}(K,\Omega)=-4\pi K\sum_\alpha e_\alpha \int \frac{G_\alpha}{\Omega-Kv}dv, \label{Delta-c}
\end{equation}
Next, a substitution of  Eq. \eqref{phiF-n2-l0}  back into Eq. \eqref{fF-alpha-n2-l0} results a slow beat wave solution given by
\begin{equation}
\tilde{f}^{(2)}_{\alpha,0}\doteq k^2\left[-\frac{{\cal{W}}(K,\Omega)}{\Delta^{(c)}(K,\Omega)} \frac{K}{\Omega-Kv}G_\alpha(v)-\frac{e_\alpha}{m_\alpha}\frac{K}{\Omega-Kv}I_\alpha(v)\right] H(K,\Omega). \label{fF-alpha-n2-l0-final}  
\end{equation}  
\section{Modes with $n=3,~l=1$ and the NLS equation}\label{appendix6}
We can now use the expressions for the lower-order quantities so obtained to determine the components  for $n=3$ and $l=1$. Thus,  we have from Eq. \eqref{vlasov1}
\begin{eqnarray}
&& f^{(3)}_{\alpha,1}\doteq -k \frac{G_\alpha}{\omega-kv}\phi^{(3)}_1+ i\frac{G_\alpha}{\omega-kv}\frac{\partial}{\partial \eta}\phi^{(2)}_1+i\frac{G_\alpha}{\omega-kv}\frac{\partial}{\partial \zeta}\phi^{(1)}_1 \notag\\
&& -i\frac{1}{(\omega-kv)^2}\left\lbrace-kG_\alpha\left( \frac{\partial}{\partial \sigma}+ v\frac{\partial}{\partial \eta}\right)\phi^{(2)}_1-i \left( \frac{\partial}{\partial \sigma}+ v\frac{\partial}{\partial \eta}\right)\left[ \frac{\partial}{\partial \sigma}  f^{(1)}_{\alpha,1} +v\frac{\partial}{\partial \eta}  f^{(1)}_{\alpha,1}- G_\alpha
\frac{\partial}{\partial \eta} \phi^{(1)}_1\right]    \right\rbrace \notag\\
&& +ik\frac{G_\alpha}{(\omega-kv)^2}v \frac{\partial}{\partial \zeta}\phi^{(1)}_1 +2 \frac{e_\alpha}{m_\alpha} \frac{k^2}{\omega-kv}\frac{\partial}{\partial v}\left\lbrace \frac{G_\alpha}{\omega-kv}\right\rbrace \phi^{(1)}_{-1}\phi^{(2)}_2 \notag\\
&&-\frac{e_\alpha}{m_\alpha} \frac{k^2}{\omega-kv}\frac{\partial}{\partial v}\left\lbrace \frac{G_\alpha}{\omega-kv}\phi^{(2)}_2 -\frac{e_\alpha}{2m_\alpha} \frac{k}{\omega-kv}\frac{\partial}{\partial v}\left(  \frac{G_\alpha}{\omega-kv}\right) \phi^{(1)}_{1}\phi^{(1)}_1\right\rbrace \phi^{(1)}_{-1} -\frac{e_\alpha}{m_\alpha} \frac{k}{\omega-kv}\frac{\partial}{\partial v}f^{(2)}_{\alpha,0}\phi^{(1)}_{1}. \label{f-alpha-n3-l1}
\end{eqnarray} 
This expression for $f^{(3)}_{\alpha,1}$ is then substituted in Eq. \eqref{poisson1} for $n=3,~l=1$ to obtain the following equation
\begin{eqnarray}
&& -ik4 \pi \sum_\alpha \int\frac{G_\alpha}{(\omega-kv)^2} dv \lambda \frac{\partial}{\partial \zeta} \phi^{(1)}_1 -\frac{\partial^2}{\partial \eta^2} \phi^{(1)}_1\notag\\
&& +4 \pi \sum_\alpha e_\alpha \int\frac{1}{(\omega-kv)^2}\left\lbrace \frac{\partial}{\partial \sigma}+ v\frac{\partial}{\partial \eta} \right\rbrace \left\lbrace\frac{\partial}{\partial \sigma}  f^{(1)}_{\alpha,1} +v\frac{\partial}{\partial \eta}  f^{(1)}_{\alpha,1}- G_\alpha
\frac{\partial}{\partial \eta} \phi^{(1)}_1 \right\rbrace dv\notag\\ 
&&-k^2 4 \pi\sum_\alpha \frac{e^2_\alpha}{m_\alpha}\int\frac{1}{\omega-kv}\frac{\partial}{\partial v}\left\lbrace \frac{G_\alpha}{\omega-kv}\right\rbrace dv \phi^{(1)}_{-1}\phi^{(2)}_2 \notag\\
&& -\frac{k^3}{2}4\pi\sum_\alpha \frac{e^3_\alpha}{m^2_\alpha}\int\frac{1}{\omega-kv}\frac{\partial}{\partial v}\left\lbrace\frac{1}{\omega-kv}\frac{\partial}{\partial v} \left( \frac{G_\alpha}{\omega-kv}\right) \right\rbrace dv |\phi^{(1)}_1|^2 \phi^{(1)}_1\notag\\
&&+K4 \pi\sum_\alpha \frac{e^2_\alpha}{m_\alpha}\int\frac{1}{\omega-kv}\frac{\partial}{\partial v}f^{(2)}_{\alpha,0}dv\phi^{(1)}_{1} + i\frac{\theta(p)}{2\epsilon^2}4 \pi \sum_\alpha e_\alpha \int_\Gamma\frac{G_\alpha}{\omega-kv} dvk \phi^{(1)}_1=0, \label{phi-n3-l1}
\end{eqnarray}
where $\theta(p)$ is unity for $p=0$ and vanishes otherwise, and $\Gamma$ denotes the path of integration around $v=\omega/k$ in the anticlockwise sense. Note that  this integral results from the term
$\int_R f^{(1)}_{\alpha,1}dv-\int_C f^{(1)}_{\alpha,1}dv$ representing the Landau damping term.  

Now, substituting the   expression \eqref{fF-alpha-n2-l0-final} into the coefficient of $\phi^{(1)}_1$ of the sixth term of Eq. \eqref{phi-n3-l1} we obtain the term as  
\begin{eqnarray}
 K4\pi \sum_\alpha \frac{e_\alpha^2}{m_\alpha}\int \frac{1}{\omega-kv}\frac{\partial}{\partial v} f^{(2)}_{\alpha,0}dv=&&k^4 \frac{1}{(2\pi)^2}\int d\Omega \int dKH(K,\Omega) \exp[i(k\eta-\Omega \sigma)]\notag\\
&&\times\left[{{{\cal{W}}(K,\Omega)}^2/\Delta^{(c)}(K,\Omega) +{\cal{C}} (K,\Omega)} \right], \label{sixth-term-phi-n3-l1}
\end{eqnarray}
where 
\begin{equation}
{\cal{C}}(K,\Omega)=4\pi\sum_\alpha \frac{{e_\alpha}^3}{m^2_\alpha}\int_c \frac{K}{{(\omega-kv)^2}{(\Omega-kv)}} I_{\alpha}(v,K,\Omega) dv. \label{C-cal}
\end{equation}
By deforming the contour  $C$ to the Landau contour with real $\Omega$ we have
\begin{equation}
\frac{1}{\Omega-kv+i\nu}=\frac{1}{\Omega-kv}-i\pi \frac{1}{|K|}\delta\left( v- \frac{\Omega}{K}\right). \label{54 eq}
\end{equation}
Thus, using the relation  \eqref{H-K-Omega} for $H(K,\Omega)$, the  functions defined by  Eqs. \eqref{W-cal}, \eqref{Delta-c} and \eqref{C-cal} can, respectively, be expressed as follows:
\begin{equation}
{\cal{W}}(K,\Omega)= -W(k,\omega;\lambda)-iU(k,\omega;\lambda) \frac{K}{|K|}, \label{W-cal-final}
\end{equation}
with
\begin{equation}
W(k,\omega;\lambda)=4\pi \sum \frac{e^2_\alpha}{m_\alpha}\int \frac{1}{(\omega-kv)^2}\frac{G_\alpha}{v-\lambda}dv, \label{W-k-omega-lamb}
\end{equation}
\begin{equation}
U(k,\omega; \lambda)=4\pi^2 \sum_\alpha \frac{e_\alpha ^2}{m_\alpha} \frac{G_\alpha (\lambda)}{(\omega-k\lambda)^2}. \label{U-k-omega}
\end{equation}
Similarly, 
\begin{equation}
{\cal{C}}(K,\Omega)=C(k,\omega; \lambda)-iD(k,\omega ; \lambda)\frac{K}{|K|}, \label{C-cal-final}
\end{equation}
with
\begin{equation}
C(k,\omega; \lambda)=-4\pi\sum_\alpha \frac{e_\alpha^3}{m_\alpha^2}\int \frac{1}{(\omega-kv)^2} \frac{I_\alpha(v)}{v-\lambda}dv, \label{C-k-omega}
\end{equation}
\begin{equation}
D(k,\omega; \lambda)=4\pi^2\sum_\alpha \frac{e_\alpha^3}{m_\alpha^2}\frac{I_\alpha(\lambda)}{(\omega-k\lambda)^2}.\label{D-k-omega}
\end{equation}
Also, we have
\begin{equation}
\Delta^{(c)}(K,\Omega)=\Delta(\lambda)+i\Gamma(\lambda)\frac{K}{|K|}, \label{Delta-c-final}
\end{equation}
where
\begin{equation}
\Delta(\lambda)=4\pi \sum_\alpha e_\alpha \int \frac{G_\alpha(v)}{v-\lambda}dv, \label{Delta-lambda}
\end{equation}
\begin{equation}
\Gamma(\lambda)=4\pi^2\sum_\alpha e_\alpha G_\alpha(\lambda). \label{Gamma-lambda}
\end{equation}
We note that the group velocity $\lambda=\partial \omega/\partial k$ is independent of $K$ and $\Omega$ and we have the relation
\begin{equation}
\frac{1}{(2\pi)^2}\int d\Omega \int dK \exp i(K\eta-\Omega\sigma) \frac{K}{|K|}H(K,\Omega)= i \frac{{\cal{P}}}{\pi}\int d\xi' \frac{|\phi^{(1)}_1(\xi',\lambda)|^2}{\xi-\xi'}, \label{64 eq}
\end{equation}
in terms of the constrained coordinate  \eqref{xi-transf}. Thus,  we  finally obtain the term given by Eq. \eqref{sixth-term-phi-n3-l1} as
\begin{equation}
k4\pi\sum_\alpha \frac{e^2_\alpha}{m_\alpha}\int \frac{1}{\omega-kv}\frac{\partial}{\partial v}f^{(2)}_{\alpha,0}dv \phi^{(1)}_1 
=k\Theta(k,\omega)|\phi^{(1)}_1|^2\phi^{(1)}_1 +k\Phi(k,\omega)\frac{{\cal{P}}}{\pi}\int \frac{|\phi^{(1)}_1(\xi',\tau)|^2}{\xi-\xi'}d\xi'\phi^{(1)}_1, \label{sixth-term-phi-n3-l1-final}
\end{equation}
where the symbols used are 
\begin{equation}
\Theta(k,\omega)=k^3\left[ \frac{\Delta}{\Delta^2+\Gamma^2}(W^2-U^2)+2 \frac{\Gamma}{\Delta^2+\Gamma^2}WU+C\right] , \label{Theta-k-omega}
\end{equation}
\begin{equation}
\Phi(k,\omega)=k^3\left[\frac{\Gamma}{\Delta^2+\Gamma^2}(W^2-U^2-2 \frac{\Delta}{\Delta^2+\Gamma^2}WU+D\right]. \label{Phi-k-omega}
\end{equation}
Here, the functions $U,~D$ and $\Gamma$ represent contributions of the resonant particles having the group velocity of the wave envelope in EP plasmas. Now, the third term in Eq. \eqref{phi-n3-l1} can be reduced by   eliminating $f^{(1)}_{\alpha,1}$ with the help of  Eq. \eqref{f-alpha-l1} as
\begin{equation}
-4\pi \sum_\alpha e_\alpha \int \frac{v-\lambda}{(\omega-kv)^2} G_\alpha dv \frac{\partial^2}{\partial \eta^2}\phi^{(1)}_1 -k4\pi \sum_\alpha e_\alpha \int \frac{(v-\lambda)^2}{(\omega-kv)^3} G_\alpha dv \frac{\partial^2}{\partial \eta^2}\phi^{(1)}_1, \label{third-term-phi-n3-l1}
\end{equation}
in which the first term of it cancels the second term of Eq. \eqref{phi-n3-l1} by means of the expression of the group velocity \eqref{lambda}. Furthermore, inspecting on the first term of Eq. \eqref{phi-n3-l1} we find that the  derivative $\lambda{\partial}/{\partial \xi}$   can be transformed into the derivative ${\partial}/{\partial \tau}$ by redefining the variable $\zeta=\lambda\tau$.
Finally, substitutions of Eqs. \eqref{phi2-n2-l2} and \eqref{sixth-term-phi-n3-l1-final} into Eq. \eqref{phi-n3-l1} give    
\begin{equation}
i\frac{\partial\phi}{\partial \tau}+P \frac{\partial^2\phi}{\partial \xi^2}+Q |\phi|^2 \phi +\frac{R }{\pi}{\cal P}\int\frac{ |\phi(\xi',\tau)|^2}{\xi-\xi'} d\xi' \phi+iS\phi=0. \label{nls-eqn}
\end{equation}



\begin{thebibliography}{50}
\bibitem{misner1973} W. Misner, K. S. Thorne, and J. A. Wheeler, {\it Gravitation} (Freeman, San Francisco, 1973), p. 763; G. W. Gibbons, S. W. Hawking, and S. Siklos, {\it The Very Early Universe} (Cambridge
University Press, Cambridge, UK, 1983).
\bibitem{lightman1982} A. P. Lightman, Astrophys. J. {\bf253}, 842 (1982); M. L. Burns and R. V. E.
Lovelace, {\it ibid.} {\bf262}, 87 (1982); A. P. Lightman and A. A. Zdziarski, {\it ibid.} {\bf 319}, 643 (1987); M. Y. Yu, P. K. Shukla, and L. Stenflo, {\it ibid.} {\bf309}, L63 (1986).
\bibitem{blandford1977} R. D. Blandford  and R. L. Znajek,  {\it Electromagnetic extraction of energy from
Kerr black holes} Mon. Not. R. Astron. Soc. {\bf 179}, 433  (1977).
\bibitem{goldreich1969} P. Goldreich and W. H. Julian, {\it Pulsar electrodynamics} Astrophys. J. {\bf157},
869  (1969).
\bibitem{wardle1998} J. F. C. Wardle  {\it et al.}, {\it Electron-positron jets associated with the quasar 3C279} Nature {\bf 395}, 457 (1998).
\bibitem{begelman1984} M. C. Begelman, R. D. Blandford, and M. D. Rees, Rev. Mod. Phys. {\bf56}, 255 (1984); H. R. Miller and P. J. Witta, in {\it Active Galactic Nuclei} (Springer, Berlin, 1987), p. 202.
\bibitem{orsoz1997} J. R. Orsoz, R. A. Remillard, C. D. Bailyn, and J. E. McClintock, Astrophys. J. {\bf478}, L83 (1997).
\bibitem{sarri2015} G. Sarri {\it et al.}, {\it Generation of neutral and high-density electron–positron pair plasmas in the laboratory} Nat. Commun. {\bf 6}, 6747 (2015).
\bibitem{surko1989} C. M. Surko, M. Leventhal, and A. Passner, Phys. Rev. Lett. {\bf62}, 901 (1989); R. G. Greaves and C. M. Surko, {\it ibid.} {\bf75}, 3846 (1995).
\bibitem{boehmer1995} H. Boehmer, M. Adams, and N. Rynn, Phys. Plasmas {\bf2}, 4369 (1995).
\bibitem{oohara2003}  W. Oohara  and R. Hatakeyama, Phys. Rev. Lett. {\bf 91}, 205005 (2003).
\bibitem{oohara2005} W. Oohara, D. Date, and R. Hatakeyama, Phys. Rev. Lett. {\bf95}, 175003
(2005).
\bibitem{jao2014} C.-S. Jao and L.-N. Hau, Phys. Rev. E {\bf 89}, 053104 (2014).
\bibitem{saberian2013} E. Saberian and A. Esfandyari-Kalejahi, Phys. Rev. E \textbf{87}, 053112 (2013). 
\bibitem{misra2004} A. P. Misra and A. Roy Chowdhury, Phys. Rev. E {\bf70}, 058401 (2004).
\bibitem{liu2014} D. Liu {\it et al.}, Phys. Plasmas {\bf21}, 022108 (2014).
\bibitem{mofiz1990} U. A. Mofiz, Phys. Rev. A {\bf42}, 960 (1990).
\bibitem{asenjo2012} F. A. Asenjo {et al.}, Phys. Rev. E {\bf85}, 046406 (2012).
\bibitem{javan2012} N. S. Javan, Phys. Plasmas {\bf19}, 122107 (2012).
\bibitem{ott1969} E. Ott and R. N. Sudan, Phys. Fluids {\bf12}, 2388 (1969); {\it ibid} {\bf13}, 1432 (1970).
\bibitem{misra2014} A. Barman and A. P. Misra, Phys. Plasmas {\bf 21}, 073708 (2014).
\bibitem{misra2015} A. P. Misra and A. Barman, Phys. Plasmas {\bf22}, 073708 (2015).
\bibitem{brodin2015} G. Brodin, J. Zamanian, and J. T. Mendonca, Phys. Scr. {\bf90}, 068020 (2015).
\bibitem{montgomery1968} M. D. Montgomery, S. J. Bame, and A. J. Hundhausen, J. Geophys. Res. {\bf73}, 4999 (1968);  M. Maksimovic, V. Pierrard, and P. Riley, Geophys. Res. Lett. {\bf24}, 1151 (1997); I. Zouganelis, J. Geophys. Res. {\bf113}, A08111 (2008).
\bibitem{liu1994} J. M. Liu  {\it et al.}, Phys. Rev. Lett. {\bf72} (1994) 2717.
\bibitem{pierrard2010} V. Pierrard  and M. Lazar   Solar Phys. {\bf267} 153 (2010) and references therein.
\bibitem{renyi1955} A. Renyi, Acta Math. Acad. Sci. Hungar.{\bf 6}, 285 (1955).
\bibitem{tsallis1988} C. Tsallis, J. Stat. Phys. {\bf 52}, 479 (1988).
\bibitem{caruso2008} F. Caruso and C. Tsallis, Phys. Rev. E {\bf78}, 021102 (2008).
\bibitem{nobre2011} F. D. Nobre, M. A. Rego-Monteiro, and C. Tsallis, Phys. Rev. Lett. {\bf106}, 140601 (2011).
\bibitem{guo2013} S. Guo, L. Mei, and A. Sun, Annals. Phys. {\bf332}, 38 (2013).
\bibitem{leubner2002} M. P. Leubner, Astrophys. Space Sci. {\bf282}, 573 (2002).
\bibitem{landau1946} L. Landau, J. Phys. (USSR) {\bf10}, 25 (1946).
\bibitem{malmberg1964} J. H. Malmberg and C. B. Wharton, Phys. Rev. Lett. {\bf13}, 184 (1964).
\bibitem{ikezi1971} H. Ikezi and Y. Kiwamoto, Phys. Rev. Lett. {\bf27}, 718 (1971).
\bibitem{ichikawa1974} Y. H. Ichikawa, Suppl. Progress Th. Phys. \textbf{55}, 212 (1974).
\bibitem{silva1998} R. Silva, Jr., A. R. Plastino, and J. A. S. Lima, Phys. Lett. A \textbf{249},
401 (1998).
\bibitem{curado1999} E. M. F. Curado, Braz. J. Phys. {\bf29}, 36 (1999); S. Abe, Physica
(Amsterdam) {\bf 269A}, 403 (1999).
\bibitem{lima2001} J. A. S. Lima, R. Silva, and A. R. Plastino, Phys. Rev. Lett. {\bf86},
2938 (2001).
\bibitem{taniuti1969} T. Taniuti and N. Yajima, J. Math. Phys. {\bf10}, 1369 (1969).
\bibitem{fedele2002} R. Fedele, H. Schamel, and P. K. Shukla, Phys. Scr. {\bf T98}, 18 (2002);  R. Fedele, Phys. Scr. {\bf65}, 502 (2002);  R. Fedele and H. Schamel, Eur. Phys. J. B {\bf27}, 313 (2002).







\end{thebibliography}
\end{document}